\documentclass[a4paper,fleqn]{cas-sc}

\usepackage[authoryear,longnamesfirst]{natbib}

\usepackage{stmaryrd} 
\usepackage{empheq}   
\usepackage[ruled,linesnumbered]{algorithm2e} 
\usepackage{eurosym} 
\PassOptionsToPackage{obeyspaces}{url}
\usepackage{subfig} 

\def\tsc#1{\csdef{#1}{\textsc{\lowercase{#1}}\xspace}}
\tsc{WGM}
\tsc{QE}
\tsc{EP}
\tsc{PMS}
\tsc{BEC}
\tsc{DE}

\begin{document}
\let\WriteBookmarks\relax
\def\floatpagepagefraction{1}
\def\textpagefraction{.001}

\shorttitle{Integrated Charging Scheduling and Operational Control for an Electric Bus Network}

\shortauthors{R. Lacombe et~al.}

\title [mode = title]{Integrated Charging Scheduling and Operational Control for an Electric Bus Network}

\tnotemark[1]

\tnotetext[1]{The paper has been supported by the Swedish Energy Agency through the project “Operational Network Energy Management for Electrified Buses” (46365-1) and IRIS with Transport Area of Advance (Chalmers University of Technology).}


%
\author[1]{Rémi Lacombe}
\cormark[1]


\ead{lacombe@chalmers.se}



\affiliation[1]{organization={Department of Electrical Engineering},
    addressline={Chalmers University of Technology},
    city={Gothenburg},
    country={Sweden}}

\author[1]{Nikolce Murgovski}
\ead{nikolce.murgovski@chalmers.se}

\author[2]{Sébastien Gros}
\ead{sebastien.gros@ntnu.no}

\affiliation[2]{organization={Department of Engineering Cybernetics},
    addressline={Norwegian University of Science and Technology (NTNU)},
    city={Trondheim},
    country={Norway}}

\author[1]{Balázs Kulcsár}
\ead{kulcsar@chalmers.se}

\cortext[cor1]{Corresponding author}



\begin{abstract}
    The last few years have seen the massive deployment of electric buses in many existing transit networks. However, the planning and operation of an electric bus system differ from that of a bus system with conventional vehicles, and some key problems have not yet been studied in the literature. In this work, we address the integrated operational control and charging scheduling problem for a network of electric buses with a limited opportunity charging capacity. We propose a hierarchical control framework to solve this problem, where the charging and operational decisions are taken jointly by solving a mixed-integer linear program in the high-level control layer. Since this optimization problem might become very large as more bus lines are considered, we propose to apply Lagrangian relaxation in such a way as to exploit the structure of the problem and enable a decomposition into independent subproblems. A local search heuristic is then deployed in order to generate good feasible solutions to the original problem. This entire Lagrangian heuristic procedure is shown to scale much better on transit networks with an increasing number of bus lines than trying to solve the original problem with an off-the-shelf solver. The proposed procedure is then tested in the high-fidelity microscopic traffic environment Vissim on a bus network constructed from an openly available dataset of the city of Chicago. The results show the benefits of combining the charging scheduling decisions together with the real-time operational control of the vehicles as the proposed control framework manages to achieve both a better level of service and lower charging costs over control baselines with predetermined charging schedules.
\end{abstract}



\begin{keywords}
Electric buses \sep Charging scheduling \sep Operational planning \sep Battery and charging \sep Optimal control \sep Lagrangian relaxation \sep Dual decomposition
\end{keywords}

\maketitle

\section{Introduction} \label{sec:1}

The decarbonization of the transport sector represents a major challenge due to the high reliance of the sector on fossil fuels. Unlike other sectors like the industry or the energy sector, global emissions from the transport sector have continued to rise steadily in both OECD and non-OECD countries, where they accounted for 30$\%$ of all $\text{CO}_2$ emissions in the former and 16$\%$ in the latter in 2016 \citep{ITF}. In an attempt to break the oil dependency of the transport sector, electric vehicles are now being massively deployed on the roads of the world, with record sales in the last few years despite the coronavirus pandemic \citep{IEA}. In particular, electric buses have emerged as a solution to not only curb greenhouse gas emissions of public transport but also to decrease air and noise pollution in urban centers \citep{bloomberg}. This explains why electric bus fleets are rapidly being adopted by municipalities around the world, several of which have already committed to reach specific electrification targets in the current decade \citep{pelletier}.

As bus fleet electrification is a relatively recent phenomenon, so too is the scientific literature on the planning process of an electric bus network. \cite{perumal} give an up-to-date overview of this new area of research, where they outline the main problems studied by researchers so far. The electric bus planning process is usually decomposed into a \textit{strategic}, a \textit{tactical}, and an \textit{operational} stage, depending on the time frame of the problem considered. For instance, strategic planning has to do with long-term decisions on charging infrastructure and bus fleet investment and on charging infrastructure placement. \cite{xylia} run a large case study to study the cost-effectiveness of electrifying the transit system in the city of Stockholm, considering both regular conductive charging and inductive charging. For conductive charging, they conclude that chargers should ideally be located at the major public transport hubs of the city. Similarly, a model that selects the best charging station locations is presented by \cite{an}. In that paper, the authors consider an uncertain charging demand, which leads them to formulate a stochastic integer program that they solve with a procedure based on Lagrangian relaxation.

At the tactical stage of the planning process, shorter-term problems such as the electric vehicle scheduling problem are considered. For a given set of bus trips, \cite{wen} propose to minimize the total number of buses deployed by solving an optimization problem in which each individual trip is assigned to a specific vehicle. The optimization problem in question is formulated as a mixed-integer linear program (MILP), and the authors deploy an adaptive large neighborhood search heuristic to solve it. Another key problem at the tactical stage of the electric bus planning process, which is of particular importance for this paper, is the charging scheduling problem. Broadly speaking, charging scheduling is about planning when and where vehicles should charge their batteries. In the electric bus literature, charging scheduling usually takes the form of an optimization problem where charging costs should be minimized, often by considering a time-of-use electricity pricing scheme \citep{perumal}. Charging scheduling for the transit network in Davis, California is formulated as an MILP by \cite{wang}, and the results of their analysis show that electrifying this network would be more economical for the transit agency than continuing to run diesel buses. Fast charging is considered in that paper, but the charging durations are assumed to be the same for all vehicles and are set at a fixed value for the entire day. By contrast, partial charging is allowed in the charging scheduling problem formulated by \cite{he}. In that work, it is assumed that buses are free to access fast chargers simultaneously, without any restrictions on the number of available chargers. The problem is first assembled as an optimal control problem and later discretized to a more tractable linear program that is then solved with an off-the-shelf solver. A more recent work manages to handle both partial charging and constraints on the number of chargers available simultaneously \citep{huang}. A solution method using Lagrangian relaxation is proposed by the authors to tackle the resulting integer linear program, and their method is shown to outperform an off-the-shelf solver for several artificial bus networks.

Finally, the operational stage of the electric bus planning process deals with real-time operational decisions when vehicles are in operation. Most of the work carried out so far at the operational stage of the planning process has been done for conventional buses and without considering the specific needs of electric buses, as explained in the recent review paper \cite{gkiotsalitis3}. Although given the short time frame of these real-time operational control problems, the difference between conventional and electric buses might not always be critical. Whether it is for conventional or electric vehicles though, bus lines are systems that are unstable in nature since any delay in service tends to be amplified in a positive feedback loop \citep{newell}. In practice, this can make it hard to follow the predefined schedules set during the tactical stage and might ideally require schedules to be reworked to fit current traffic conditions in some cases. Transit agencies may resort to many different intervention strategies for the operational control of their vehicles. These strategies are usually divided into station control strategies, the most popular of which is bus holding \citep{daganzo2}, and inter-station control strategies, the most popular of which is speed control \citep{varga1}. Both bus holding and speed control have their own merits and may complement each other well, which is why we consider a combination of both in this article.

The different stages of the electric bus planning process are often treated separately in the literature due to their inherent complexity \citep{perumal}, but some researchers have recently started proposing \textit{integrated} frameworks that include problems from different stages of the planning process. For example, \cite{dan} develop a framework in which decisions on charger placement, vehicle scheduling, and charging scheduling are taken jointly for the transit network of Seattle. However, the real-time operational control aspect has mostly been left out of these integrated approaches so far. We think that this oversight of operational aspects is problematic since transit systems are stochastic, and it is not uncommon in practice for unplanned disturbances to cause service delays that can severely disrupt any planned schedule designed at the tactical stage. Guaranteeing operational robustness is in fact identified as a main future research direction in the review paper \cite{perumal}. This is especially relevant for electric buses due to their potentially short driving range and their potentially limited access to the charging infrastructure, though this also depends on the charging technology used \citep{häll}. In this article, we consider daytime opportunity charging, which consists in partially recharging bus batteries during daytime operations with fast chargers located at certain bus stops or nearby depots. Due to the generally high costs of high-power chargers, the charging capacity tends to be limited in opportunity charging. In addition, buses tend to have smaller batteries since they are not meant to complete a full day of bus operations with a single charge. In light of all these considerations, an integrated framework combining charging scheduling and real-time operational control is developed in what follows.

This paper presents a full-fledged control framework based on the prototype that was first introduced in \cite{lacombe4}. The following changes and improvements have been made over this reference:
\begin{itemize}
    \item The bus network model used in the high-level control layer has been extended with several new key features to better capture the real dynamics of a transit system. In particular, the model now includes time-of-use electricity pricing, vehicle mass and capacity constraints, as well as a more detailed energy consumption model.
    \item A scalable procedure based on a Lagrangian relaxation and a dual decomposition is proposed in this paper to find good solutions to the optimization problems that must be solved in the high-level control layer. As such, the proposed control framework is not reliant on solving possibly large MILPs to optimality with an off-the-shelf solver, as is the case in the work cited above, something which can not reasonably be done on more than very small networks.
    \item In contrast to the synthetic simulations presented in \cite{lacombe4}, this paper offers an extensive case study of a small bus network in Chicago run in the microscopic traffic environment Vissim. By simulating all vehicles and all traffic intersections explicitly, Vissim is able to capture the real operations of a transit system much more faithfully than would be possible in synthetic simulations. Emerging phenomena, such as the formation of traffic queues due to bus operations during rush hours, can now be accounted for in the simulation environment.
\end{itemize}

The main contribution of this paper is to combine together two aspects of the planning process of an electric bus network which are traditionally treated separately, namely the charging scheduling aspect and the real-time operational aspect. The benefit of doing so is that operational aspects are fundamental in determining how well any predefined schedule can be followed in practice, and they must therefore be used to update the schedule in real-time in order for robust performance to be achieved. Moreover, a detailed model of an electric bus network is presented in this work, one which includes travel time commands on individual inter-stop links, passenger queues at stops, capacity constraints, rush hour traffic, piecewise linear energy consumption functions for each individual inter-stop link, partial charging, limited charging capacity, and time-of-use electricity pricing. The Lagrangian relaxation and dual decomposition procedure that are applied to split the original problem into independent subproblems make the proposed control framework scale well on larger transit systems. The general idea behind this decomposition could in principle be transferred to similar transit planning problems where bus lines are weakly coupled together, as a way to improve the scalability of these problems.

The article is organized as follows. The model equations of the electric bus network considered are presented in Section \ref{sec:2}. Then, in Section \ref{sec:3}, it is shown how the operational bus network control and charging scheduling problem can be assembled as an MILP. In order to solve this problem efficiently, a decomposition procedure based on Lagrangian relaxation is presented in Section \ref{sec:4}. Simulation results are then shown and analyzed in Section \ref{sec:5}, before the paper closes on some concluding remarks in Section \ref{sec:6}.

\section{Modeling} \label{sec:2}

The control framework presented in this article to solve the operational bus network control and charging scheduling problem is hierarchical and consists of a centralized high-level layer and decentralized low-level layers for all vehicles. Due to the limited capacity of the shared charging infrastructure, the charging scheduling decisions must be taken in common for all bus lines. Therefore, all the charging scheduling aspects are included in the centralized high-level control layer, together with long-term operational decisions. These high-level decisions are then used as references in the low-level control layers, where the real-time operational control of the vehicles is carried out.

Concerning the operational control aspect, we consider that the main control strategy deployed by the transit agency is speed control. This speed control strategy is also complemented with some bus holding, but it is assumed that vehicles can only be held at a single location: the bus terminal at the end of their route. Indeed, applying bus holding as a control strategy is conditional on having enough space at the bus stops, or control points, to hold vehicles without disrupting bus operations nor the surrounding traffic. In dense urban areas, where holding space is scarce, this might only be doable at larger bus stations or dedicated bus depots.

In this work, the focus is mostly being put on the design of the high-level control layer, and it is assumed that the individual low-level control layers consist of simple proportional speed controllers. Note, however, that more advanced controllers, such as one generating energy-efficient speed profiles as proposed in \cite{lacombe2}, could in principle be deployed depending on the available computing capacity of the vehicles.

In the rest of this section, the deterministic bus network model used in the high-level control layer is presented.

\subsection{Assumptions and notation}

We consider a transit system where all bus lines only have a single stop in common, that will be referred to as the shared terminal, or simply terminal, hereafter. Apart from this shared terminal, bus routes are assumed to not have any overlap. We further assume that all chargers are either located at this single shared bus terminal, or at a single shared bus depot that vehicles can only access from the shared bus terminal. These assumptions are crucial to assemble a tractable version of the operational bus network control and charging scheduling problem as they greatly reduce the complexity of the model by removing some of the coupling between bus lines. We believe, however, that the problem formulated in this article could still be used as a tractable approximation of the bus network control problem in the case where some of these assumptions are not verified, such as when some bus routes overlap. Note also that modeling bus charging at intermediary bus stops could be added easily as it would only require slight changes to the proposed model.

\begin{figure}[b]
\begin{center}
\includegraphics[width=0.6\textwidth]{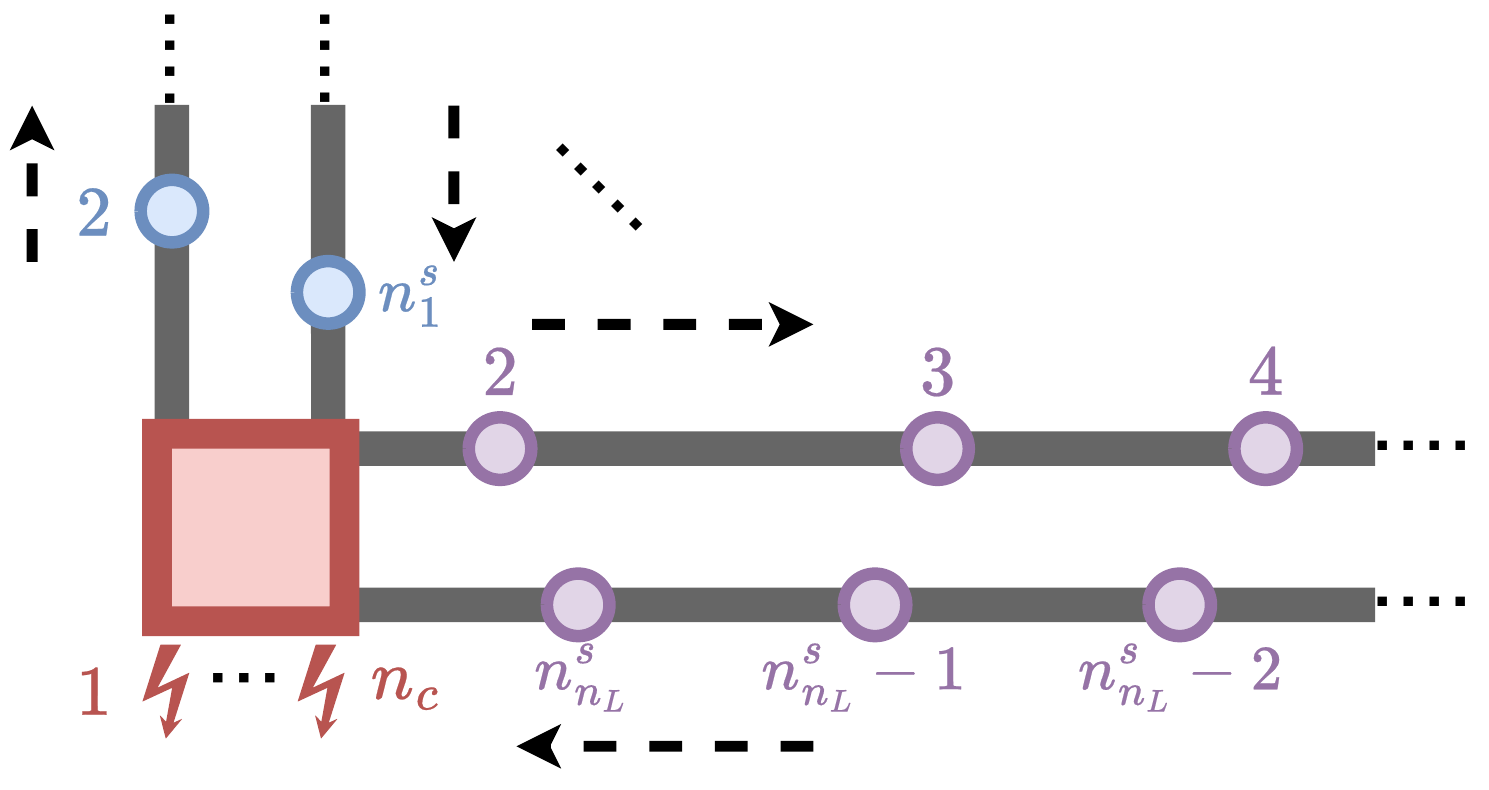}
\caption{Representation of the bus network structure considered. The shared bus terminal is displayed as a big red square. The chargers are shown in red close to the terminal. Only the bus routes with index $1$ and index $n_L$ are shown. A few of the bus stops on these routes are represented as circles. Black arrows indicate the direction of driving.}
\label{fig:bus_network}
\end{center}
\end{figure}

The bus network considered is assumed to consist of $n_L$ bus lines. Let ${\mathcal{L}=\{1,...,n_L\}}$ be the set of line indices. For any bus line ${l\in\mathcal{L}}$, the sets of bus and stop indices are defined as ${\mathcal{I}_l=\{1,...,n^b_l\}}$ and ${\mathcal{J}_l=\{1,...,n^s_l\}}$, respectively, and the target headway of the line is noted $H_l$. Without loss of generality, the shared terminal is the stop with index 1 in $\mathcal{J}_l$, $\forall l \in \mathcal{L}$. Similarly, by assuming that $n_c$ chargers are located at the terminal, or at a shared bus depot, the set of charger indices is noted $\mathcal{O}={\{1,...,n_c\}}$. In the rest of this paper, index $i$ will often be used to refer to a bus, index $j$ to a bus stop, and index $o$ to a charger. Figure \ref{fig:bus_network} provides an illustration of the bus network.

In order for the bus network model assembled in this section to be practically useful, the system is only modeled over a given finite time horizon. In some cases, this finite horizon can be long enough for several full trips to be completed by every bus considered. Therefore, we now introduce a visit index $k$ in case of repeated visits of the same bus to the same stops. Without loss of generality, let ${\mathcal{K}_{l,i,j}=\{1,...,n^v_{l,i,j}\}}$ be the set of indices for visits of bus $i$ of line $l$ at stop $j$, where $n^v_{l,i,j}$ is the number of visits that bus $i$ of line $l$ is expected to make to the stop with index $j$ over the time horizon considered. Note that it is assumed that overtaking is not allowed and that vehicles may not be redeployed on a different bus line upon reaching the terminal.

In what follows, all the constraints and state equations presented hold for any ${l\in\mathcal{L}}$, ${i\in\mathcal{I}_l}$, ${j\in\mathcal{J}_l}$, ${o\in\mathcal{O}}$, and $k\in\mathcal{K}_{l,i,j}$, unless stated otherwise.

\subsection{Bus dynamics}

First, the state variable $t^k_{l,i,j}$ is introduced as the arrival time of bus $i$ of line $l$ at stop $j$ for its $k$-th visit at that stop. Before writing the corresponding state equation, some more constraints and variables need to be introduced.

We start with the no-overtaking assumption, which can be enforced between consecutive bus visits as:
\begin{equation} \label{eq:no_overtaking}
    t_{l,i,j}^k-t_{l,i-1,j}^k \geq 0,
\end{equation}
where an index $i=0$ refers to the bus of line $l$ with index $n_l^b$.

The travel time of bus $i$ of line $l$ on the $j$-th link of the route, between the stops with index $j$ and $j+1$, is noted $\tau_{l,i,j}^k$ and is bounded above and below as:
\begin{equation} \label{eq:time_bounds}
    T^{\text{min},k}_{l,j} \leq \tau_{l,i,j}^k \leq T^{\text{max},k}_{l,j}.
\end{equation}
The upper travel time bound represents the travel time when traveling at a given minimum acceptable speed, while the lower travel time bound comes from whatever may limit the maximum speed of buses in operations, such as speed limits or the average influence of the surrounding traffic and traffic lights. Note that these bounds may vary with the visit index $k$ in order to model traffic conditions that might change at different times of the day.

A standard passenger model is used to model the passenger accumulation at stops between subsequent bus visits. As is common practice in the literature, passenger arrivals at each stop are modeled as a homogeneous Poisson process \citep{hans}. The mean passenger arrival rate is noted $\lambda_{l,j}^k$ for stop $j$ of line $l$. Note that this term is also dependent on the visit index $k$ in order to model the variations in passenger volumes over the day. As an example, the expected number of boarding passengers that bus $i$ of line $l$ finds at stop $j$ during its $k$-th visit is proportional to the amount of time that has passed since the preceding bus visited that stop, and is given by: ${\lambda_{l,j}^k(t_{l,i,j}^k-t_{l,i-1,j}^k)}$.

\begin{table*}[!h]
\renewcommand{\arraystretch}{1.1}
\caption{Summary of all indices and variables used in Sections \ref{sec:2} and \ref{sec:3}.}
\label{tb:dv}
\centering
{\begin{tabular}{|lllc|} \hline
& Symbol & Description & Interval \\ \hline
Indices & $l$ & Bus line index & $\mathcal{L}$ \\
& $i$ & Bus index & $\mathcal{I}_l$ \\
& $j$ & Bus stop or inter-stop link index & $\mathcal{J}_l$ \\
& $o$ & Charger index & $\mathcal{O}$ \\
& $k$ & Stop visit index & $\mathcal{K}_{l,i,j}$ \\ \hline
State variables & $t_{l,i,j}^k$ & Arrival time at stop & $\mathbb{R}_+$ \\
& $m_{l,i,j}^k$ & Mass & $\mathbb{R}_+$ \\
& $\sigma_{l,i,j}^k$ & State-of-charge & $[0,1]$ \\ \hline
Control inputs & $\tau_{l,i,j}^k$ & Inter-stop travel time & $\mathbb{R}_+$ \\
& $w_{l,i}^{\text{before},k}$ & Holding time before charging & $\mathbb{R}_+$ \\
& $w_{l,i}^{\text{after},k}$ & Holding time after charging & $\mathbb{R}_+$ \\
& $c_{l,i}^k$ & Charging time & $\mathbb{R}_+$ \\
& $b_{l,i,o}^k$ & Charging decision & $\{0,1\}$ \\
& $\psi_{\{l,i,k\},\{l',i',k'\}}$ & Charging order & $\{0,1\}$ \\ \hline
Slack variables & $E_{l,i,j}^k$ & Energy consumption & $\mathbb{R}_+$ \\
& $r_{l,i,j}^k$ & Refused passengers & $\mathbb{R}_+$ \\
& $\eta_{l,i,j}^k$ & Positive headway deviation at stop & $\mathbb{R}_+$ \\
& $\nu_{l,i}$ & Terminal cost & $[0,1]$ \\ \hline
\end{tabular}}
\end{table*}

\begin{table*}[!h]
\renewcommand{\arraystretch}{1.1}
\caption{Summary of all modeling parameters used in Section \ref{sec:2}.}
\label{tb:parameters_modeling}
\centering
{\begin{tabular}{|ll|} \hline
Symbol & Description \\ \hline
$n_L$ & Number of bus lines \\
$n_c$ & Number of chargers \\
$n_l^b$ & Number of buses on line $l$ \\
$n_l^s$ & Number of bus stops on line $l$ \\
$H_l$ & Target headway on line $l$ \\
$T_{l,j}^{\text{min},k}$ & Lower travel time bound  \\
$T_{l,j}^{\text{max},k}$ & Upper travel time bound \\ 
$\lambda_{l,j}^k$ & Passenger arrival rate \\
$\xi_{l,j}$ & Passenger alighting proportion \\
$m_\text{pax}$ & Passenger mass \\
$m_\text{cap}$ & Bus capacity \\
$d_\text{pax}$ & Passenger boarding time \\
$d_\text{char}$ & Time delay for charging \\
$P_\text{char}$ & Charging power \\
$Q_{l,i}$ & Battery capacity \\
$E_0$ & Energy to travel to the depot \\
$\sigma_l^\text{min}$ & Minimum state-of-charge allowed \\
$M$ & Big-M coefficient \\ \hline
\end{tabular}}
\end{table*}

Next, another state variable is introduced: the mass of bus $i$ of line $l$ when arriving at stop $j$ for its $k$-th visit at that stop is noted $m_{l,i,j}^k$. Since the bus mass is constant between consecutive stops, $m_{l,i,j}^k$ also denotes the bus mass when leaving the previous stop. The state equation for the mass is:
\begin{equation} \label{eq:mass_dynamics}
    m_{l,i,j+1}^k = (1-\xi_{l,j})m_{l,i,j}^k + m_{\text{pax}}\lambda_{l,j}^k(t_{l,i,j}^k-t_{l,i-1,j}^k) - m_{\text{pax}}r_{l,i,j}^k,
\end{equation}
where $\xi_{l,j}$ is the average proportion of onboard passengers who alight at that stop and $m_{\text{pax}}$ is the average passenger mass. In this equation, $r_{l,i,j}^k$ is a decision variable which denotes the number of passengers waiting at the stop to which boarding in that bus is refused.

Refusing waiting passengers from boarding a bus is not meant to be part of the control strategy applied by the transit agency in this work. This new variable is only used here as a way to model the limited capacity of the vehicles and not as an alternative control strategy, and will thus only be nonzero when there is a risk that the bus capacity can be exceeded. The way in which this can be done is explained later in the article. For now, it can be observed that the number of passengers to which boarding is refused cannot exceed the expected number of passengers waiting at a stop, and so:
\begin{equation} \label{eq:refused_bound}
    0 \leq r_{l,i,j}^k \leq \lambda_{l,j}^k(t_{l,i,j}^k-t_{l,i-1,j}^k).
\end{equation}

With the notation introduced so far, the bus capacity constraint is simply:
\begin{equation} \label{eq:mass_bound}
     m_{l,i,j}^k \leq m_{\text{cap}},
\end{equation}
where all vehicles are assumed to have the same mass capacity $m_{\text{cap}}$. Observe that no lower bound on the mass is needed since the mass cannot be decreased by more than the total mass of passengers currently onboard in \eqref{eq:mass_dynamics}.

Let us now consider a terminal visit. The shared terminal plays a special role compared with the other bus stops since this is where buses can charge their batteries, and also where bus holding is allowed. Each binary charging decision variable $b_{l,i,o}^k$ takes value 1 if bus $i$ of line $l$ uses charger $o$ during its $k$-th visit at the terminal, and 0 otherwise. Each bus can use at most one charger during each terminal visit, which is materialized by the constraint:
\begin{equation} \label{eq:sum_b}
    \sum_{o=1}^{n_c}b_{l,i,o}^k \leq 1, \hfill k\in\mathcal{K}_{l,i,1}.
\end{equation}

Let $c_{l,i}^k$ be the charging time corresponding to that terminal visit. The charging time must be zero in the event of a zero charging decision, which can be enforced through:
\begin{equation} \label{eq:charging_time_constraint}
    0 \leq c_{l,i}^k \leq M \sum_{o=1}^{n_c}b_{l,i,o}^k, \hfill k\in\mathcal{K}_{l,i,1}
\end{equation}
where $M$ is a large positive number used in the so-called big-M method \citep{richards}. Note that the charging time is allowed to take any positive value here.

We are now ready to write the dynamics of the time state variable:
\begin{empheq}[left={t_{l,i,j+1}^k=\empheqlbrace}]{align}
    & t_{l,i,j}^k + w_{l,i}^{\text{before},k} + 2d_{\text{char}}\sum_{o=1}^{n_c}b_{l,i,o}^k + c_{l,i}^k + w_{l,i}^{\text{after},k} + \tau_{l,i,j}^k, \hspace{3.3cm} && j = 1, \label{eq:time_dynamics1} \\
    & t_{l,i,j}^k + \tau_{l,i,j}^k + d_{\text{pax}}\lambda_{l,j}^k(t_{l,i,j}^k-t_{l,i-1,j}^k) - d_{\text{pax}}r_{l,i,j}^k, && j\in\llbracket2,n^s_l\rrbracket, \label{eq:time_dynamics2}
\end{empheq}
where stop indices $1$ and $n^s_l+1$ are used interchangeably to denote the terminal, and where the average time needed for each passenger boarding is noted $d_{\text{pax}}$. The expression given for a terminal visit in \eqref{eq:time_dynamics1} includes bus holding through the continuous variables $w_{l,i}^{\text{before},k}$ and $w_{l,i}^{\text{after},k}$, which enables holding before and after using a charger, respectively. In case the decision has been made for the bus to use one charger, $d_{\text{char}}$ denotes the time delay to start and end the charging operation. Note that $d_{\text{char}}$ may be used just as well to model a scenario where the chargers are located directly at the end terminal as one in which they are located at an adjacent bus depot.

At other stops, the evolution of the time state variable is determined by the inter-link travel times and the dwell times at stops, as expressed in \eqref{eq:time_dynamics2}. Note that it is assumed in this expression that the dwell time of buses at stops results from the passenger exchange operation and is determined by the boarding passengers only. This is a frequent modeling assumption which is motivated by the fact that the boarding operation often takes longer to complete than the alighting operation \citep{hans}.

Finally, we consider that the entire passenger exchange operation at the terminal is carried out upon arrival of a vehicle there. This means that only passengers that are already waiting when a bus arrives at the terminal can board it. Passengers arriving later, when the bus is held at the terminal or using a charger, are not allowed to board it. The minimum holding time when arriving at the terminal must then be at least equal to the time needed to complete the passenger exchange:\begin{equation} \label{eq:holding_bound}
    w_{l,i}^{\text{before},k} \geq d_{\text{pax}}\lambda_{l,1}^k(t_{l,i,1}^k-t_{l,i-1,1}^k) - d_{\text{pax}}r_{l,i,1}^k, \hfill k\in\mathcal{K}_{l,i,1}.
\end{equation}
This assumption greatly simplifies the equations by avoiding the need to resort to queuing models. Though this may seem restrictive, a simple way to improve the modeling accuracy is to lower the passenger arrival rate at the terminal to account for the newly arrived passengers directly boarding any bus waiting at the terminal.

\subsection{Bus charging}

Let $\sigma_{l,i,j}^k$ be a state variable denoting the state-of-charge of bus $i$ of line $l$ when arriving at stop $j$ for its $k$-th visit at that stop. The state-of-charge dynamics are modeled as:
\begin{empheq}[left={\sigma_{l,i,j+1}^k=\empheqlbrace}]{align}
    & \sigma_{l,i,j}^k + \frac{P_{\text{char}}}{Q_{l,i}}c_{l,i}^k - 2\frac{E_0}{Q_{l,i}}\sum_{o=1}^{n_c}b_{l,i,o}^k - \frac{E_{l,i,j}^k}{Q_{l,i}}, \hspace{5.8cm} && j = 1, \label{eq:SoC_dynamics1} \\
    & \sigma_{l,i,j}^k - \frac{E_{l,i,j}^k}{Q_{l,i}}, && j\in\llbracket2,n^s_l\rrbracket, \label{eq:SoC_dynamics2}
\end{empheq}
where $P_{\text{char}}$ is the charging power, $Q_{l,i}$ is the battery capacity of bus $i$ of line $l$, and $E_{l,i,j}^k$ is a variable representing the energy consumption of the bus when traveling on the $j$-th inter-stop link of the route. As mentioned previously, the chargers might be located at a separate bus depot, in which case any vehicle planning to charge would have to consume an additional fixed quantity of energy $E_0$ to travel between the terminal and the depot. If the chargers are located at the bus terminal, we simply have ${E_0=0}$.

In this work, the energy consumption is modeled as a piecewise linear function of both the travel time and the mass. The motivation for this choice is that it ultimately makes it possible to formulate a linear optimization problem, as discussed in the next section, whereas higher-order energy consumption models would lead to a class of more complex optimization problems. It is assumed that the same number $n_q$ of linear pieces are used for all inter-stop links. It is considered that each linear piece corresponds to a feasible travel time interval, and that the travel time breakpoints are noted $T^{k,1}_{l,j},T^{k,2}_{l,j},...,T^{k,n_q+1}_{l,j}$ in increasing order, where $T^{k,1}_{l,j}=T^{\text{min},k}_{l,j}$ and $T^{k,n_q+1}_{l,j} =T^{\text{max},k}_{l,j}$. With these modeling choices, the energy consumption is given by:
\begin{equation} \label{eq:energy_consumption}
    E_{l,i,j}^k = a_{1,l,j}^q\tau_{l,i,j}^k + a_{2,l,j}^qm_{l,i,j}^k + a_{3,l,j}^q, \quad q\in\llbracket1,n_q\rrbracket \hspace{0.2cm} \text{s.t.} \hspace{0.2cm} T^{k,q}_{l,j} \leq \tau_{l,i,j}^k \leq T^{k,q+1}_{l,j},
\end{equation}
where the linear function coefficients for the piece with index $q$ are noted $a_{1,l,j}^q$, $a_{2,l,j}^q$, and $a_{3,l,j}^q$.

Next, we consider state-of-charge constraints. We start by ensuring that all state-of-charge values are between 0 and 1:
\begin{equation} \label{eq:SoC_bounds}
    0 \leq \sigma_{l,i,j}^k \leq 1.
\end{equation}

In addition, it is assumed that buses must have a high enough state-of-charge when leaving the terminal to avoid running out of battery before the next charging opportunity. Looking at equation \eqref{eq:SoC_dynamics1}, this condition can be enforced as:
\begin{equation} \label{eq:SoC_feasibility}
    \sigma_{l,i,1}^k + \frac{P_{\text{char}}}{Q_{l,i}}c_{l,i}^k - 2\frac{E_0}{Q_{l,i}}\sum_{o=1}^{n_c}b_{l,i,o}^k \geq \sigma_l^{\text{min}}, \hfill k\in\mathcal{K}_{l,i,1},
\end{equation}
where $\sigma_l^{\text{min}}$ is the minimum state-of-charge value allowed when leaving the terminal. Note that this minimum value is line-specific since the different lengths and overall structure of the bus routes might result in different energy needs for completing one full trip.

It now remains to model the limited capacity of the charging infrastructure. As mentioned earlier, it is assumed that only $n_c$ high-power chargers are available to buses for daytime opportunity charging. Recall that charging durations are not limited in this work and can in principle take any positive values, as long as it does not violate the state-of-charge constraints. Therefore, the constraints that must be assembled now should mainly aim to ensure that no charger can be used by more than a single bus at any given time, and will be referred to as the \textit{charger exclusion constraints} hereafter.

For the sake of clarity, the following notation is first introduced to denote the time at which a bus begins charging:
\begin{equation} \label{eq:t_char}
    t^{\text{char},k}_{l,i} = t_{l,i,1}^k + w_{l,i}^{\text{before},k} + d_{\text{char}}, \hfill k\in\mathcal{K}_{l,i,1}.
\end{equation}

Then, for any pair of terminal visits ${\{l,i,k\}}$ and ${\{l',i',k'\}}$, with ${\{l,i,k\}\in\mathcal{L}\times\mathcal{I}_l\times\mathcal{K}_{l,i,1}}$ and $\{l',i',k'\}\in\mathcal{L}\times\mathcal{I}_{l'}\times\mathcal{K}_{l',i',1}$, the charger exclusion constraints can be written as:
\begin{equation} \label{eq:boolean}
    (t^{\text{char},k}_{l,i} + c_{l,i}^{k} \leq t^{\text{char},k'}_{l',i'}) \oplus (t^{\text{char},k'}_{l',i'} + c_{l',i'}^{k'} \leq t^{\text{char},k}_{l,i}) \lor (1-b_{l,i,o}^k) \lor (1-b_{l',i',o}^{k'}), \hfill \forall o\in\mathcal{O},
\end{equation}
where the $\oplus$ symbol represents the classical \texttt{xor} logical operation. This Boolean expression encodes that, in case the same charger with index $o$ is planned to be used in both terminal visits (i.e. $b_{l,i,o}^k=1$ and $b_{l',i',o}^{k'}=1$), then exactly one of the two buses involved must stop using the charger before the other bus starts using it. In case no charging is planned during one of the visits, or in case two separate chargers are planned to be used, expression \eqref{eq:boolean} is automatically satisfied for all chargers.

\section{Bus network control problem} \label{sec:3}

The focus of this next section is to formulate the operational bus network control and charging scheduling problem to be solved in the high-level control layer, now that a bus network model has been assembled. As mentioned before, the goal is to take charging scheduling decisions for the vehicles together with long-term operational decisions over a look-ahead control horizon. For this type of operational problems, transit agencies would usually try to optimize the quality of the service provided to the passengers while minimizing their operating costs at the same time, and the problem assembled here should reflect these two objectives.

\subsection{Model reformulation}

Some of the model equations introduced in the previous section are now modified in order to be able to write the optimal control problem in the mathematical programming framework. In addition, some of the constraints are also modified with the sole purpose of improving numerical performance when solving the problem later on.

Looking at equation \eqref{eq:time_dynamics1}, one can see that the set of control inputs ${\{w_{l,i}^{\text{after},k}\}_{l\in\mathcal{L},i\in\mathcal{I}_l,k\in\mathcal{K}_{l,i,1}}}$ for holding buses after charging is not really needed since these variables do not appear anywhere else in the model. Therefore, they can simply be removed from the model, and \eqref{eq:time_dynamics1} replaced with the equivalent inequality constraint:
\begin{equation} \label{eq:time_dynamics1_mod}
     t_{l,i,j+1}^k \geq t_{l,i,j}^k + w_{l,i}^{\text{before},k} + 2d_{\text{char}}\sum_{o=1}^{n_c}b_{l,i,o}^k + c_{l,i}^k + \tau_{l,i,j}^k, \hspace{7.2cm} j = 1.
\end{equation}
Note that the same cannot be done with the control inputs
${\{w_{l,i}^{\text{before},k}\}_{l\in\mathcal{L},i\in\mathcal{I}_l,k\in\mathcal{K}_{l,i,1}}}$ since they are needed to express the times at which buses start charging in \eqref{eq:t_char}.

Given a travel time command, the piecewise linear expression used in \eqref{eq:energy_consumption} to model the energy consumption requires the identification of the right linear piece. The implicit \texttt{if} statement included in \eqref{eq:energy_consumption} needs to be reformulated in order to fit in the mathematical programming framework. One way to do so is to relax the optimization problem by expressing the equality constraint \eqref{eq:energy_consumption} as the following set of inequality constraints instead:
\begin{equation} \label{eq:energy_consumption_relax}
    E_{l,i,j}^k \geq a_{1,l,j}^q\tau_{l,i,j}^k + a_{2,l,j}^qm_{l,i,j}^k + a_{3,l,j}^q, \hfill \forall q\in\llbracket1,n_q\rrbracket.
\end{equation}
Under some conditions, this relaxation of the optimization problem can be tight, meaning that the optimal solution remains the same. Sufficient conditions for this relaxation to be tight are: (i) penalizing higher energy consumption values in the objective function and (ii) having a convex piecewise linear function in \eqref{eq:energy_consumption}. The rationale is that the energy consumption minimization aspect ensures that the lowest possible value of $E_{l,i,j}^k$ is chosen for any given values of $\tau_{l,i,j}^k$ and $m_{l,i,j}^k$, while convexity of the piecewise linear function ensures that the right piece is chosen since only the corresponding inequality constraint would be active in \eqref{eq:energy_consumption_relax}. In what follows, we assume that conditions (i) and (ii) are satisfied.

Similarly, the Boolean expression representing the charger exclusion constraints in \eqref{eq:boolean} needs to be a written as a set of constraints in which logical operators do not appear. Using the big-M method, \eqref{eq:boolean} can equivalently be expressed as:
\begin{subequations}
\begin{align}
& t^{\text{char},k}_{l,i} + c_{l,i}^k - t^{\text{char},k'}_{l',i'} \leq \big(1-b_{l,i,o}^k\big)M + \big(1-b_{l',i',o}^{k'}\big)M + \psi_{\{l,i,k\},\{l',i',k'\}} M, \label{eq:exclusion1} \\
& t^{\text{char},k'}_{l',i'} + c_{l',i'}^{k'} - t^{\text{char},k}_{l,i} \leq \big(1-b_{l,i,o}^k\big)M + \big(1-b_{l',i',o}^{k'}\big)M + \big(1-\psi_{\{l,i,k\},\{l',i',k'\}}\big)M.,\label{eq:exclusion2}
\end{align}
\end{subequations}
where $\psi_{\{l,i,k\},\{l',i',k'\}}$ is a binary variable which encodes the charging order between terminal visits ${\{l,i,k\}}$ and ${\{l',i',k'\}}$. Here again, the constraints \eqref{eq:exclusion1} and \eqref{eq:exclusion2} are only relevant if the same charger $o$ is planned to be used for both terminal visits, i.e. ${b_{l,i,o}^k=1}$ and ${b_{l',i',o}^{k'}=1}$. In that case, a value of ${\psi_{\{l,i,k\},\{l',i',k'\}}=1}$ means that the charging event of terminal visit ${\{l,i,k\}}$ takes place after that of visit ${\{l',i',k'\}}$, and vice versa for ${\psi_{\{l,i,k\},\{l',i',k'\}}=0}$.

Due to the non-overtaking assumption, the charging order for buses of the same line is already known since a bus is not allowed to charge before the bus preceding it. For any pair of terminal visits ${\{l,i,k\}}$ and ${\{l',i',k'\}}$ such that $l=l'$, the charger exclusion constraints for any charger $o$ can simply take the form:
\begin{equation} \label{eq:same_line}
    t^{\text{char},k}_{l,i} + c_{l,i}^k - t^{\text{char},k'}_{l,i'} \leq \big(1-b_{l,i,o}^k\big)M + \big(1-b_{l,i',o}^{k'}\big)M,
\end{equation}
where it is assumed that ${\{l,i,k\}}$ is the earlier visit. No additional binary variable is needed in this case, which greatly simplifies the overall problem.

The vectors gathering all state variables ${t=\{t_{l,i,j}^k\}_{l\in\mathcal{L},i\in\mathcal{I}_l,j\in\mathcal{J}_l,k\in\mathcal{K}_{l,i,j}}}$, ${m=\{m_{l,i,j}^k\}_{l\in\mathcal{L},i\in\mathcal{I}_l,j\in\mathcal{J}_l,k\in\mathcal{K}_{l,i,j}}}$, and $\sigma=\{\sigma_{l,i,j}^k\}_{l\in\mathcal{L},i\in\mathcal{I}_l,j\in\mathcal{J}_l,k\in\mathcal{K}_{l,i,j}}$ are used to assemble the state vector ${x=[t,m,\sigma]^{\top}}$. Similarly, the control input vectors ${\tau=\{\tau_{l,i,j}^k\}_{l\in\mathcal{L},i\in\mathcal{I}_l,j\in\mathcal{J}_l,k\in\mathcal{K}_{l,i,j}}}$, ${w=\{w^{\text{before},k}_{l,i}\}_{l\in\mathcal{L},i\in\mathcal{I}_l,k\in\mathcal{K}_{l,i,1}}}$, ${c=\{c^k_{l,i}\}_{l\in\mathcal{L},i\in\mathcal{I}_l,k\in\mathcal{K}_{l,i,1}}}$, ${b=\{b^k_{l,i,o}\}_{l\in\mathcal{L},i\in\mathcal{I}_l,o\in\mathcal{O},k\in\mathcal{K}_{l,i,1}}}$, and    ${\psi=\{\psi_{\{l,i,k\},\{l',i',k'\}}\}_{l\in\mathcal{L},l'\in\mathcal{L}\backslash\{l\},i\in\mathcal{I}_l,i'\in\mathcal{I}_{l'},k\in\mathcal{K}_{l,i,1},k'\in\mathcal{K}_{l',i',1}}}$ can be gathered in a control input vector $u=[\tau,w,c,b,\psi]^{\top}$. Finally, the slack variables ${E=\{E_{l,i,j}^k\}_{l\in\mathcal{L},i\in\mathcal{I}_l,j\in\mathcal{J}_l,k\in\mathcal{K}_{l,i,j}}}$ and ${r=\{r_{l,i,j}^k\}_{l\in\mathcal{L},i\in\mathcal{I}_l,j\in\mathcal{J}_l,k\in\mathcal{K}_{l,i,j}}}$ are grouped in the vector ${s=[E,r]^{\top}}$. All primal decision variables of the optimization problem are then gathered in the vector ${z=[x,u,s]^{\top}}$.

\subsection{Horizon and cost function}

The design of the control horizon on which the bus network control problem is formulated is now discussed. First, one must observe that one specific characteristic of the dynamics of a bus network is that the largest changes in the state equations of the vehicles take place at bus stops, whose spatial location is fixed and known beforehand. Modeling changes in the dynamics at predefined spatial points is nontrivial in the time domain, and usually requires some heavy adjustments of the state equations, see e.g. \cite{hult6} for the case of a single traffic intersection. Since there are many stop visits planned in the problem of interest here, we made the choice of formulating the control problem in the space domain, which is why the system equations have mostly been assembled in the space domain in the previous section.

Let us consider the optimal control problem over a finite time control horizon of length $T$. This temporal horizon must then be mapped to an equivalent spatial horizon. However, the way to carry out this mapping is not unique for the system considered since each vehicle faces different local conditions. There can for example be a denser distribution of bus stops in certain parts of the transit network, or across different bus lines. This means that all vehicles are unable to cover the exact same distance in $T$ time units, and therefore that the finite time horizon has to be mapped to a different spatial horizon for every vehicle.

Next, an economic objective function is considered for this problem, meaning that each objective term is given a monetary value. This function is meant to represent the costs to the transit agency incurred by the operation of the transit system considered. The cost function considered here is:
\begin{equation} \label{eq:objective_function}
    J(x,u,s) = \sum_{\substack{l\in\mathcal{L},i\in\mathcal{I}_l \\ j\in\mathcal{J}_l,k\in\mathcal{K}_{l,i,j}}}
    \left[p_{\text{reg}}\text{max}\big(0,t_{l,i,j}^k-t_{l,i-1,j}^k-H_l\big) + p_{\text{reject}} r_{l,i,j}^k\right] + \sum_{\substack{l\in\mathcal{L},i\in\mathcal{I}_l \\ k\in\mathcal{K}_{l,i,1}}} p_{\text{el}}^k c_{l,i}^k + \sum_{\substack{l\in\mathcal{L},i\in\mathcal{I}_l}} p_{\text{end}} \text{max}(0,\sigma_{\text{goal}}-\sigma_{l,i}^{\text{end}})
\end{equation}
and consists of four terms with different purposes:
\begin{itemize}
    \item The first term penalizes positive headways deviations, that is arrivals at stops which happen later than the target headway of the bus line $H_l$. The coefficient $p_{\text{reg}}$ applies a monetary cost to headway deviations and can be based on passengers' value of time or on fines applied to the transit operator for being late, for instance.
    \item The second term applies a large penalty $p_{\text{reject}}$ for refusing passenger boarding. This connects back to what was mentioned earlier: refusing passenger boarding is not used as a control strategy in this work and should only be seen as a convenient way to use continuous slack variables. The rationale for using these slack variables here is that additional binary variables would have otherwise been needed to model capacity constraints, in order to detect which of the bus mass variables exceed the capacity limit. A large value of $p_{\text{reject}}$ thus ensures that the boarding refusal variables are only nonzero if it is absolutely needed due to the high penalty that this incurs.
    \item The third term is simply the total charging cost on the horizon considered. Here, the electricity price $p_{\text{el}}^k$ depends on the index of the terminal visit. This is meant to model time-of-use electricity pricing: the electricity price might be different at different times of the planning horizon.
    \item The fourth term is akin to a terminal cost, and penalizes bus state-of-charge which are lower than a certain desired value $\sigma_{\text{goal}}$ at the end of the horizon. For the sake of notational convenience, we have noted $\sigma_{l,i}^{\text{end}}$ the last state-of-charge variable of bus $i$ of line $l$ on its horizon. The role of this terminal cost term is to account for the limited length of the problem horizon and to incentivize vehicles to keep enough battery for the remaining duration of their service. More information on the design of $\sigma_{\text{goal}}$ is given right after.
\end{itemize}

In a similar fashion as in the previous subsection, the maximum function in \eqref{eq:objective_function} must be reformulated in order to assemble the optimal control problem in the mathematical programming framework. This is done by adding two new sets of slack variables, which yields the equivalent expression for the cost function:
\begin{equation} \label{eq:objective_function_slack}
    J(x,u,s) = \sum_{\substack{l\in\mathcal{L},i\in\mathcal{I}_l \\ j\in\mathcal{J}_l,k\in\mathcal{K}_{l,i,j}}}
    \left[p_{\text{reg}}\eta_{l,i,j}^k + p_{\text{reject}} r_{l,i,j}^k\right] + \sum_{\substack{l\in\mathcal{L},i\in\mathcal{I}_l \\ k\in\mathcal{K}_{l,i,1}}} p_{\text{el}}^k c_{l,i}^k + \sum_{\substack{l\in\mathcal{L},i\in\mathcal{I}_l}} p_{\text{end}} \nu_{l,i}
\end{equation}
such that:
\begin{subequations}
\begin{align}
    & \eta_{l,i,j}^k \geq 0, \label{eq:slack1} \\
    & \eta_{l,i,j}^k \geq t_{l,i,j}^k-t_{l,i-1,j}^k-H_l, \label{eq:slack2} \\
    & \nu_{l,i} \geq 0, \label{eq:slack3} \\
    & \nu_{l,i} \geq \sigma_{\text{goal}}-\sigma_{l,i}^{\text{end}}. \label{eq:slack4}
\end{align}
\end{subequations}
By gathering these new slack variables in the vectors ${\eta=\{\eta_{l,i,j}^k\}_{l\in\mathcal{L},i\in\mathcal{I}_l,j\in\mathcal{J}_l,k\in\mathcal{K}_{l,i,j}}}$ and ${\nu=\{\nu_{l,i}\}_{l\in\mathcal{L},i\in\mathcal{I}_l}}$, we update the vector of slack variables to be ${s=[E,r,\eta,\nu]^{\top}}$.

\subsection{Terminal cost design} \label{sec:3.3}

Before moving on to fully assemble the operational bus network control and charging scheduling problem, the question of the terminal cost design must first be addressed. This term is meant to model that the transit system might continue to run beyond the current horizon. In the problem considered here, the main issue that may arise as a result of the finiteness of the horizon is that vehicles might skip charging for as long as they are allowed to since each charging event has a cost. We would then want to design the terminal cost in such a way as to incentivize a relatively homogeneous charging across the entire simulation, such as by penalizing low state-of-charge values early on in the simulation.

We assume that the entire simulation has a total duration $T_\text{sim}$, likely with $T<T_\text{sim}$. A typical choice for $T_\text{sim}$ in this context is one day of bus operations. Assume that all the vehicles have a certain state-of-charge $\sigma_\text{ini}$ at the start of the simulation and that they aim to have a state-of-charge of $\sigma_\text{end}$ at the end of the simulation. Electric buses would typically start their day of operations with a full battery, and try to end it with a relatively empty battery, both as a way to reduce layover time during the day and to benefit from cheaper electricity prices during overnight charging. Note that $\sigma_\text{ini}$ and $\sigma_\text{end}$ are assumed here to be the same for all vehicles of all bus lines for the sake of simplicity, but that these values could easily be adapted to generate a different $\sigma_\text{goal}$ function for each individual bus.

A straightforward way to design the terminal cost is to model $\sigma_\text{goal}$ as a linear function of the simulation time $t_\text{sim}$, that is the time at which the control problem is considered, such as:
\begin{equation} \label{eq:terminal_cost_linear}
    \sigma_\text{goal}(t_\text{sim}) = \sigma_\text{end} + \frac{T_\text{sim}-T-t_\text{sim}}{T_\text{sim}}(\sigma_\text{ini}-\sigma_\text{end}), \hfill t_\text{sim} \in [0,T_\text{sim}-T].
\end{equation}
Observe that the desired state-of-charge value at the end of the horizon is shifted by $T$ time units compared to the desired value at the current simulation time in this expression. For instance, we have that $\sigma_\text{goal}(T_\text{sim}-T)=\sigma_\text{end}$ because the end of the simulation coincides with the end of the horizon when the optimization problem is formed at time ${T_\text{sim}-T}$.

Even though the linear function in \eqref{eq:terminal_cost_linear} can be a good design choice for the terminal cost, it can be improved further. Since the hourly electricity prices are known one day in advance in our time-of-use electricity pricing model, it might also be valuable to embed this information in the terminal cost design. By anticipating the incoming variation in electricity prices, the controller can increase the charging load when prices are lower.

We now consider a piecewise linear function design for $\sigma_\text{goal}$ where each piece corresponds to one simulation hour. We further assume that the total duration of the simulation $T_\text{sim}$ consists in an integer number of hours $N$. The electricity price during the $n$-th hour of simulation is noted $p_{\text{el},n}$, with ${n\in\llbracket1,N\rrbracket}$, and the average electricity price over the entire simulation is noted $\overline{p}_\text{el}$. The main idea in the proposed terminal cost design is to attribute non-uniform weights to the desired hourly charging amounts. More specifically, the weights in question steer the speed of decrease of the desired state-of-charge, and are expressed as:
\begin{equation} \label{eq:sigma_weights}
    \omega_n = \frac{1+\epsilon\big(p_{\text{el},n}-\overline{p}_\text{el}\big)}{N}, \hfill n\in\llbracket1,N\rrbracket,
\end{equation}
where $\epsilon$ is a positive coefficient that weighs how much the desired charging pattern should differ from a uniform hourly charging. Note that $\epsilon$ should be chosen small enough to guarantee that all weights are non-negative, and it can be seen that the weights sum to 1 in that case.

The desired state-of-charge at the end of each new simulation hour is then given by the recursion:
\begin{subequations} \label{eq:sigma_recursion}
\begin{align}
    & \sigma_0 = \sigma_\text{ini}, \\
    & \sigma_n = \sigma_{n-1} - \omega_n(\sigma_\text{ini}-\sigma_\text{end}), \hspace{8.9cm} n\in\llbracket1,N\rrbracket,
\end{align}
\end{subequations}
where $\sigma_n$ is the desired state-of-charge right at the end of the $n$-th hour of simulation. Therefore, lower electricity prices lead to lower weights which in turn lead to a slower decrease of the desired state-of-charge: vehicles are incentivized to keep a high state-of-charge and therefore to charge their battery while prices are low.

The expression for the desired vehicle state-of-charge at any moment of the simulation is then:
\begin{equation} \label{eq:terminal_desired_cost}
    \sigma_\text{des}(t_\text{sim}) = \frac{\sigma_n-\sigma_{n-1}}{3600}t_\text{sim} + \sigma_n - \frac{\sigma_n-\sigma_{n-1}}{3600}T_n, \quad n\in\llbracket1,N\rrbracket \hspace{0.2cm} \text{s.t.} \hspace{0.2cm} T_{n-1} \leq t_\text{sim} \leq T_n,
\end{equation}
where $\{T_n\}_{n\in\llbracket0,N\rrbracket}$ is the set of simulation times at which electricity prices change. They are assumed to coincide with the simulations hours here, such that $T_0=0$, $T_N=T_\text{sim}$, and that $\sigma_n$ is the desired state-of-charge at time $T_n$.

As before, the actual value of $\sigma_\text{goal}$ is obtained by shifting the desired state-of-charge by $T$ time units, since the terminal cost applies to the desired state-of-charge at the end of the horizon. Therefore, we have:
\begin{equation} \label{eq:terminal_cost}
    \sigma_\text{goal}(t_\text{sim}) = \sigma_\text{des}(t_\text{sim}+T), \hfill t_\text{sim} \in [0,T_\text{sim}-T].
\end{equation}

Since time-of-use pricing is included in our simulation framework, the expression \eqref{eq:terminal_cost} is used for the terminal cost. Note, however, that one could use the linear version \eqref{eq:terminal_cost_linear} instead in an environment with constant electricity prices.

\subsection{Problem formulation}

We are now ready to assemble the operational bus network control and charging scheduling problem:
\begin{subequations} \label{eq:MILP}
\begin{align}
    \underset{x,u,s}{\text{min}} \quad & J(x,u,s) = \sum_{\substack{l\in\mathcal{L},i\in\mathcal{I}_l \\ j\in\mathcal{J}_l,k\in\mathcal{K}_{l,i,j}}}
    \left[p_{\text{reg}}\eta_{l,i,j}^k + p_{\text{reject}} r_{l,i,j}^k\right] + \sum_{\substack{l\in\mathcal{L},i\in\mathcal{I}_l \\ k\in\mathcal{K}_{l,i,1}}} p_{\text{el}}^k c_{l,i}^k + \sum_{\substack{l\in\mathcal{L},i\in\mathcal{I}_l}} p_{\text{end}} \nu_{l,i} \\
    \text{s.t.} \quad & \: \forall l \in \mathcal{L}, \forall i \in \mathcal{I}_l: \nonumber \\
    & \: \eqref{eq:no_overtaking}, \eqref{eq:time_bounds}, \eqref{eq:mass_dynamics}, \eqref{eq:refused_bound}, \eqref{eq:mass_bound}, \eqref{eq:time_dynamics2}, \eqref{eq:SoC_dynamics1}, \eqref{eq:SoC_dynamics2}, \eqref{eq:SoC_bounds},\eqref{eq:energy_consumption_relax}, \eqref{eq:slack1}-\eqref{eq:slack2},\hspace{2.3cm} \forall j \in \mathcal{J}_l, \forall k \in \mathcal{K}_{l,i,j}, \\
    & \: \eqref{eq:sum_b}, \eqref{eq:charging_time_constraint}, \eqref{eq:holding_bound},\eqref{eq:SoC_feasibility}, \eqref{eq:time_dynamics1_mod}, \eqref{eq:slack3}-\eqref{eq:slack4}, \\
    & \: b_{l,i,o}^k \in \{0,1\}, \hspace{8.7cm} \forall k \in \mathcal{K}_{l,i,1}, \forall o \in \mathcal{O}, \label{eq:binary_in_MILP} \\
    & \: \forall\{l,i,k\}\in\mathcal{L}\times\mathcal{I}_l\times\mathcal{K}_{l,i,1},\forall\{l',i',k'\}\in\mathcal{L}\times\mathcal{I}_{l'}\times\mathcal{K}_{l',i',1},\forall o \in \mathcal{O}: \nonumber \\
    & \: \eqref{eq:exclusion1}-\eqref{eq:exclusion2}, \psi_{\{l,i,k\},\{l',i',k'\}}\in\{0,1\}, \hspace{7.3cm} l\neq l', \label{eq:exclusion_in_MILP} \\ 
    & \: \eqref{eq:same_line}, \hspace{12.0cm} l=l'.
\end{align}
\end{subequations}

This optimization problem is an MILP since both the objective function and all the constraints are linear functions, and since both continuous and binary variables are present. However, mixed-integer problems are notoriously difficult to solve \citep{richards}. While the problem \eqref{eq:MILP} can in principle be solved directly with traditional open-source or commercial MILP solvers, this might not be doable in practice for more than small instance of the problem. In the next section, we make the argument that there exists a more efficient way to solve this problem by leveraging its particular structure and decomposing it into simpler subproblems.

\section{Decomposition method} \label{sec:4}

Being an NP-complete problem \citep{richards}, the MILP \eqref{eq:MILP} scales poorly when adding decision variables, such as when considering a network with more bus lines, more chargers available, etc. In order for the problem to be tractable even for larger bus networks, we propose a decomposition procedure based on a Lagrangian relaxation.

The rationale for choosing a decomposition approach here is that the problem \eqref{eq:MILP} is highly structured, thanks to the inherent dynamics of a transit system. Indeed, each vehicle is only coupled to the vehicles of the same bus line directly preceding and following it, in a classical chain-like structure. But more importantly, there is only weak coupling between different bus lines in our formulation: only the charger exclusion constraints involve vehicles from different bus lines. Indeed, by organizing the variables in a line-by-line and bus-by-bus fashion, the constraint matrix can be shown to have a block-angular structure, i.e. the constraint matrix consists of line-specific diagonal blocks and in one wider block coming from the charger exclusion constraints and coupling together variables from different bus lines. Therefore, we propose to relax the charger exclusion constraints \eqref{eq:exclusion1}-\eqref{eq:exclusion2} for buses belonging to different lines.

Since the Lagrangian relaxation procedure consists in solving a dual problem of the original MILP \eqref{eq:MILP} \citep{gaudioso}, it might not be possible to solve \eqref{eq:MILP} to optimality when using it since there might be a nonzero duality gap, i.e. a difference between the optimal value of \eqref{eq:MILP} and of its dual version. However, note that the problem \eqref{eq:MILP} is in fact too complex to be solved to optimality in most cases, except for very small bus networks. At any rate, Lagrangian relaxation can provide lower bounds on the optimal value of \eqref{eq:MILP}. Empirical results suggest that the bounds given by the Lagrangian relaxation tend to be “extremely sharp”, i.e. very close to the optimal value, as stated by \cite{fisher}. In addition, feasible solutions of good quality can in general be generated from the solutions provided by Lagrangian relaxation. Doing so usually involves some sort of problem-dependent heuristic. Combining Lagrangian relaxation with such a heuristic to find feasible solutions to the original problem is a procedure generally known as a \textit{Lagrangian heuristic} \citep{fisher}. In the electric bus planning literature, Lagrangian heuristics have successfully been used to solve optimization problems at the strategic level by \cite{an}, and at the tactical level by \cite{huang}, where the Lagrangian heuristic developed by the authors widely outperforms a commercial solver when solving one version of the electric bus scheduling problem in the latter reference. In this section, we present a Lagrangian heuristic to solve, or at least find good feasible solutions to, the operational bus network control and charging scheduling problem \eqref{eq:MILP}.

Note that other similar decomposition methods could in principle be deployed to solve the structured MILP \eqref{eq:MILP}. In particular, we believe that applying the Dantzig-Wolfe decomposition method could be a viable alternative \citep{conejo}, but this is beyond the scope of this paper. Since all coupling across bus lines comes from the charger exclusion constraints, we believe that other popular decomposition methods that are more geared towards variable coupling, such as Bender decomposition, would not be adapted here \citep{conejo}.

\subsection{Formulation of the Lagrangian dual problem}

In Lagrangian relaxation, a so-called \textit{Lagrangian dual problem} of the original problem \eqref{eq:MILP} is formed and then solved \citep{gaudioso}. To form this dual problem, we start by assembling the Lagrangian function of \eqref{eq:MILP}. First, we write the scalar product of the relaxed constraints with their associated Lagrange multipliers:
\vspace{-0.5cm}
\begin{subequations} \label{eq:relaxed_constraints}
\begin{align}
    & \sum_{\substack{\{l,i,k\}\in\mathcal{L}\times\mathcal{I}_l\times\mathcal{K}_{l,i,1}\\\{l',i',k'\}\in\mathcal{L}\times\mathcal{I}_{l'}\times\mathcal{K}_{l',i',1}\\o\in\mathcal{O},l\neq l'}}
    \begin{aligned}
        & \\
        & \beta^o_{1\{l,i,k\},\{l',i',k'\}}\big(t^{\text{char},k}_{l,i} + c_{l,i}^k - t^{\text{char},k'}_{l',i'} + \big(b_{l,i,o}^k + b_{l',i',o}^{k'} - \psi_{\{l,i,k\},\{l',i',k'\}} - 2\big)M\big) \\
        & + \beta^o_{2\{l,i,k\},\{l',i',k'\}}\big(t^{\text{char},k'}_{l',i'} + c_{l',i'}^{k'} - t^{\text{char},k}_{l,i} + \big(b_{l,i,o}^k + b_{l',i',o}^{k'} + \psi_{\{l,i,k\},\{l',i',k'\}} - 3\big)M\big)
    \end{aligned} \nonumber \\
    & = \beta^{\top}(A_{\text{relax}}z-b_{\text{relax}}), \tag{\ref{eq:relaxed_constraints}}
\end{align}
\end{subequations}
where $\beta=\big\{\beta^o_{1\{l,i,k\},\{l',i',k'\}},\beta^o_{2\{l,i,k\},\{l',i',k'\}}\big\}_{\{l,i,k\}\in\mathcal{L}\times\mathcal{I}_l\times\mathcal{K}_{l,i,1},\{l',i',k'\}\in\mathcal{L}\times\mathcal{I}_{l'}\times\mathcal{K}_{l',i',1},o\in\mathcal{O},l\neq l'}$ is a vector containing all the Lagrange multipliers corresponding to the relaxed constraints \eqref{eq:exclusion1}-\eqref{eq:exclusion2}, and $A_{\text{relax}}$ and $b_{\text{relax}}$ gather the coefficients of all those constraints. In what follows, $\beta$ is simply referred to as the vector of Lagrange multipliers for the sake of simplicity.

Having introduced these notations, the Lagrangian function can now be written as:
\begin{equation} \label{eq:lagrange_function}
    L(z,\beta) = J(z) + \beta^{\top}(A_{\text{relax}}z-b_{\text{relax}}).
\end{equation}

The Lagrangian dual function is then constructed by minimizing the Lagrangian function for a given vector of Lagrange multipliers $\beta$, thus forming the optimization problem:
\begin{subequations} \label{eq:dual_function}
\begin{align}
    h(\beta) = \: & \underset{z}{\text{min}} \quad L(z,\beta) & \\
    & \: \text{s.t.} \quad \forall l \in \mathcal{L}, \forall i \in \mathcal{I}_l: \nonumber \\
    & \hspace{0.9cm} \eqref{eq:no_overtaking}, \eqref{eq:time_bounds}, \eqref{eq:mass_dynamics}, \eqref{eq:refused_bound}, \eqref{eq:mass_bound}, \eqref{eq:time_dynamics2}, \eqref{eq:SoC_dynamics1}, \eqref{eq:SoC_dynamics2}, \eqref{eq:SoC_bounds},\eqref{eq:energy_consumption_relax}, \eqref{eq:slack1}-\eqref{eq:slack2},\hspace{1.4cm} \forall j \in \mathcal{J}_l, \forall k \in \mathcal{K}_{l,i,j}, \\
    & \hspace{0.9cm} \eqref{eq:sum_b}, \eqref{eq:charging_time_constraint}, \eqref{eq:holding_bound},\eqref{eq:SoC_feasibility}, \eqref{eq:time_dynamics1_mod}, \eqref{eq:slack3}-\eqref{eq:slack4}, \\
    & \hspace{0.9cm} \eqref{eq:same_line}, \hspace{7cm} \forall i'\in \mathcal{I}_l, \forall \{k,k'\}\in \mathcal{K}_{l,i,1}^2, \forall o \in \mathcal{O}, \\
    & \hspace{0.9cm} b_{l,i,o}^k \in \{0,1\}, \hspace{7.8cm} \forall k \in \mathcal{K}_{l,i,1}, \forall o \in \mathcal{O}, \\
    & \hspace{0.9cm} \psi_{\{l,i,k\},\{l',i',k'\}}\in\{0,1\}, \hspace{3.6cm} \forall l'\in \mathcal{L}\backslash\{l\}, \forall i'\in \mathcal{I}_{l'}, \forall \{k,k'\}\in \mathcal{K}_{l,i,1}^2.
\end{align}
\end{subequations}
The constraints in \eqref{eq:dual_function} are the same as in \eqref{eq:MILP}, with the key exception of the charger exclusion constraints, which have been relaxed. Consequently, the optimization problem \eqref{eq:dual_function} is separable across bus lines. New notations are now introduced to make this more apparent.

The primal variables can be reorganized in a line-by-line fashion as $z=\left[\{z_l\}_{l\in\mathcal{L}},\psi\right]^{\top}$. In this expression, the vector $z_l$ contains all primal variables in $z$ which only have a single bus line subscript $l$. Only the charging order decisions variables in $\psi$ have two bus line subscripts, which is why they are represented separately.

Note that, in the optimization problem \eqref{eq:dual_function}, the charging order variables contained in $\psi$ only appear in the objective function as linear terms, since $\beta$ is a fixed set of parameters in this problem. Therefore, the value of the binary variables in $\psi$ can be obtained directly from the sign of the associated linear coefficients as:
\begin{equation} \label{eq:psi}
    \psi_{\{l,i,k\},\{l',i',k'\}} = \text{max}\Big(\text{sign}\Big(\sum_{o\in\mathcal{O}}\beta^o_{1\{l,i,k\},\{l',i',k'\}}-\beta^o_{2\{l,i,k\},\{l',i',k'\}}\Big),0\Big),
\end{equation}
for all pairs of terminal visits ${\{l,i,k\}}$ and ${\{l',i',k'\}}$ such that ${l\neq l'}$. In this relaxed optimization problem, the variables in $\psi$ can thus simply be fixed by having access to the vector of Lagrange multipliers $\beta$. Without loss of generality, we adopt the notation $\psi(\beta)$ to express this dependency in the rest of this section.

Observe that the objective function \eqref{eq:objective_function_slack} is separable across bus lines, as is the relaxed constraints term of the Lagrangian function \eqref{eq:relaxed_constraints} for a fixed vector $\beta$. Consequently, the Lagrangian function itself is separable and can be decomposed as:
\begin{equation} \label{eq:lagrange_function_split}
    L(z,\beta) = \sum_{l\in\mathcal{L}} L_l(z_l,\beta) + R(\beta)
\end{equation}
where $L_l$ groups all the terms in \eqref{eq:lagrange_function} that contain a unique bus line subscript $l$, and where the fixed residual terms coming from \eqref{eq:relaxed_constraints} are gathered in $R$.

With the help of these notations, the optimization problem in \eqref{eq:dual_function} can be decomposed into $n_L$ independent subproblems, where the subproblem for bus line $l$, with $l\in\mathcal{L}$, can be written as:
\begin{subequations} \label{eq:subproblem}
\begin{align}
    h_l(\beta) = \: & \underset{z_l}{\text{min}} \quad L_l(z_l,\beta) \label{eq:subproblem1} \\
    & \: \text{s.t.} \quad \forall i \in \mathcal{I}_l: \nonumber \\
    & \hspace{0.9cm} \eqref{eq:no_overtaking}, \eqref{eq:time_bounds}, \eqref{eq:mass_dynamics}, \eqref{eq:refused_bound}, \eqref{eq:mass_bound}, \eqref{eq:time_dynamics2}, \eqref{eq:SoC_dynamics1}, \eqref{eq:SoC_dynamics2}, \eqref{eq:SoC_bounds},\eqref{eq:energy_consumption_relax}, \eqref{eq:slack1}-\eqref{eq:slack2},\hspace{1.25cm} \forall j \in \mathcal{J}_l, \forall k \in \mathcal{K}_{l,i,j}, \label{eq:subproblem2} \\
    & \hspace{0.9cm} \eqref{eq:sum_b}, \eqref{eq:charging_time_constraint}, \eqref{eq:holding_bound},\eqref{eq:SoC_feasibility}, \eqref{eq:time_dynamics1_mod}, \eqref{eq:slack3}-\eqref{eq:slack4}, \label{eq:subproblem3}\\
    & \hspace{0.9cm} \eqref{eq:same_line}, \hspace{6.8cm} \forall i'\in \mathcal{I}_l, \forall \{k,k'\}\in \mathcal{K}_{l,i,1}^2, \forall o \in \mathcal{O}, \\
    & \hspace{0.9cm} b_{l,i,o}^k \in \{0,1\}, \hspace{7.6cm} \forall k \in \mathcal{K}_{l,i,1}, \forall o \in \mathcal{O}, \label{eq:subproblem4}
\end{align}
\end{subequations}
for a given vector of Lagrange multipliers $\beta$. The optimal solution of each subproblem is noted $z_l(\beta)$, that is: $z_l(\beta) = \text{argmin}_{z_l} \left(L_l(z_l,\beta) | \forall i \in \mathcal{I}_l: \eqref{eq:subproblem2},\eqref{eq:subproblem3},\eqref{eq:subproblem4}\right)$. Each of the subproblems \eqref{eq:subproblem} is an MILP of smaller size than the original problem \eqref{eq:MILP}.

Consequently, the optimization problem in the Lagrangian dual function \eqref{eq:dual_function} can simply be rewritten as:
\begin{equation} \label{eq:dual_function_split}
    h(\beta) = \: \sum_{l\in\mathcal{L}} h_l(\beta) + R(\beta).
\end{equation}

Finally, the Lagrangian dual problem is:
\begin{equation} \label{eq:lagrange_dual_problem}
    h^*= \underset{\beta\geq0}{\text{max}} \: h(\beta),
\end{equation}
and provides a lower bound $h^*$ on the optimal value of the original problem \eqref{eq:MILP}, as discussed previously. However, solving \eqref{eq:lagrange_dual_problem} might prove difficult in practice, in part due to the fact that the Lagrangian dual function $h$ is not differentiable (it is in fact piecewise linear), which rules out the direct use of traditional gradient-based methods for solving the Lagrangian dual problem, as explained by \cite{fisher}. The same can be said of second-order optimization methods, which also rely on having a smooth optimization problem. Next, we present how a solution method conceptually similar to gradient descent, the \textit{subgradient algorithm}, can be used instead.

\subsection{Subgradient algorithm}

The subgradient algorithm is the most widely-used algorithm to solve the Lagrangian dual problem due to its robust convergence properties and its relative simplicity \citep{fisher}. The algorithm proceeds by constructing a series of iterates on the vector of Lagrange multipliers $\beta$ in a way similar to a traditional gradient descend algorithm, but where gradients are replaced with subgradients, until convergence to the solution of \eqref{eq:lagrange_dual_problem}. In practice, however, the algorithm is usually interrupted prematurely as it tends to display a fast initial convergence phase, followed by a much slower later convergence phase \citep{gaudioso}.

At any iteration $\kappa$ of the subgradient algorithm, the next iterate is obtained as:
\begin{equation} \label{eq:subgradient_update}
    \beta^{\kappa+1} \leftarrow \beta^{\kappa} + \alpha^\kappa(A_{\text{relax}}z(\beta^{\kappa})-b_{\text{relax}}),
\end{equation}
where $z(\beta^{\kappa})$ is the optimal solution of the optimization problem \eqref{eq:dual_function} for the vector $\beta^{\kappa}$, and $\alpha^\kappa$ is the step size.

Here, the step size is computed using the Polyak step length rule \citep{gaudioso}:
\begin{equation} \label{eq:subgradient_step_size}
    \alpha^\kappa = \frac{\theta^\kappa\big(\overline{h}-h(\beta^\kappa)\big)}{\|A_{\text{relax}}z(\beta^{\kappa})-b_{\text{relax}}\|^2}
\end{equation}
where $\theta^\kappa\in(0,2]$ and $\overline{h}$ is an upper bound on $h^*$. Finding a tight upper bound on $h^*$ is not easy and is usually done by setting $\overline{h}$ to be the cost of the best known feasible solution of the original problem \eqref{eq:MILP}. Indeed, the cost of any feasible solution of \eqref{eq:MILP} is an upper bound on $h^*$, since $h^*$ is itself a lower bound on the optimal value of \eqref{eq:MILP}. More information on how feasible solutions can be generated from the subgradient iterates is given in the next subsection.

\begin{algorithm}[b]
\SetAlgoLined
\DontPrintSemicolon
\tcp{Initialize multipliers and problem bounds}
$\beta^1 \leftarrow \beta_0, \quad \underline{h} \leftarrow -\infty, \quad \overline{h} \leftarrow \infty$ \;
\For{$\kappa=1,...,n_{\textup{iter}}$}{
 $h(\beta^\kappa) \leftarrow R(\beta^\kappa), \quad z(\beta^{\kappa}) \leftarrow []$ \;
 \tcp{Solve subproblems}
 \For{$l=1,...,n_L$}{
   $h(\beta^\kappa) \leftarrow h(\beta^\kappa) + h_l(\beta^\kappa)$ \;
   $z(\beta^{\kappa}) \leftarrow \left[z(\beta^{\kappa}),z_l(\beta^{\kappa})\right]^{\top}$ \;
   }
 $z(\beta^{\kappa}) \leftarrow \left[z(\beta^{\kappa}),\psi(\beta^{\kappa})\right]^{\top}$  \;
 \tcp{Update lower bound}
 $\underline{h} \leftarrow \textup{max}\big(\underline{h},h(\beta^\kappa)\big)$  \;
 \tcp{Get a feasible solution}
 $\left(h^\kappa_{\textup{feas}},z^\kappa_{\textup{feas}}\right) \leftarrow \textsc{SearchHeuristic}(z(\beta^{\kappa}))$ \label{alg:heuristic_line} \;
 \tcp{Update upper bound and best incumbent}
 \If{$h^\kappa_{\textup{feas}}<\overline{h}$}{
    \vspace{0.1cm}$\overline{h} \leftarrow h^\kappa_{\textup{feas}}$  \;
    $z_{\textup{feas}} \leftarrow z^\kappa_{\textup{feas}}$  \;}
 \tcp{Perform one iteration of the subgradient algorithm}
 $\alpha^\kappa \leftarrow \frac{\theta^\kappa(\overline{h}-h(\beta^\kappa))}{\|A_{\text{relax}}z(\beta^{\kappa})-b_{\text{relax}}\|^2}$ \;
 $\beta^{\kappa+1} \leftarrow \beta^\kappa + \alpha^\kappa(A_{\text{relax}}z(\beta^{\kappa})-b_{\text{relax}})$ \;
 }
 \textbf{return} $\underline{h},\overline{h},z_{\textup{feas}}$
 \caption{Overview of the Lagrangian heuristic procedure.}
\label{alg:lagrangian_heuristic}
\end{algorithm}

Algorithm \ref{alg:lagrangian_heuristic} summarizes the entire Lagrangian heuristic procedure presented so far. The initial vector of Lagrange multipliers is noted $\beta_0$, and the current upper and lower bounds on the optimal value of the original MILP  \eqref{eq:MILP} are noted $\overline{h}$ and $\underline{h}$, respectively. Here, the subgradient algorithm is terminated after a fixed number of iterations $n_{\textup{iter}}$. At each iteration of the algorithm, a new solution $z(\beta^{\kappa})$ of the minimization problem \eqref{eq:dual_function} is obtained by solving the subproblems \eqref{eq:subproblem} and computing $\psi(\beta^\kappa)$ with \eqref{eq:psi}. However, this solution is not feasible in the original problem \eqref{eq:MILP} in general since there is no guarantee that it satisfies the relaxed constraints. This infeasible solution can still be used as a starting point by a local search heuristic in order to construct a feasible solution of \eqref{eq:MILP}. It is assumed that such a routine is used in Line \ref{alg:heuristic_line} of Algorithm \ref{alg:lagrangian_heuristic} to generate a feasible solution $z^\kappa_{\textup{feas}}$ with a cost $h^\kappa_{\textup{feas}}$ at each iteration. One such heuristic is presented in the next subsection. The vector of Lagrange multipliers is then updated with the subgradient method. The algorithm finally returns the best upper and lower bound on the optimal value of \eqref{eq:MILP}, which are used to compute the optimality gap, as well as the best feasible solution found $z_{\textup{feas}}$.

\begin{figure}[b]
\centering
\subfloat[Infeasible solution $z(\beta^\kappa)$.]
{\includegraphics[width=\textwidth]{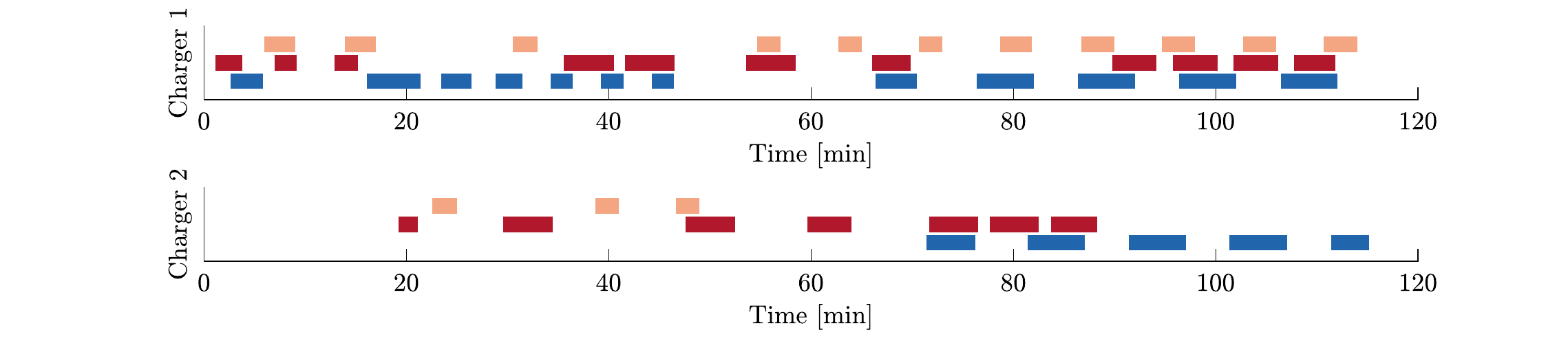} \label{fig:heuristic_infeasible}}%
\vfil
\subfloat[Basic heuristic.]
{\includegraphics[width=\textwidth]{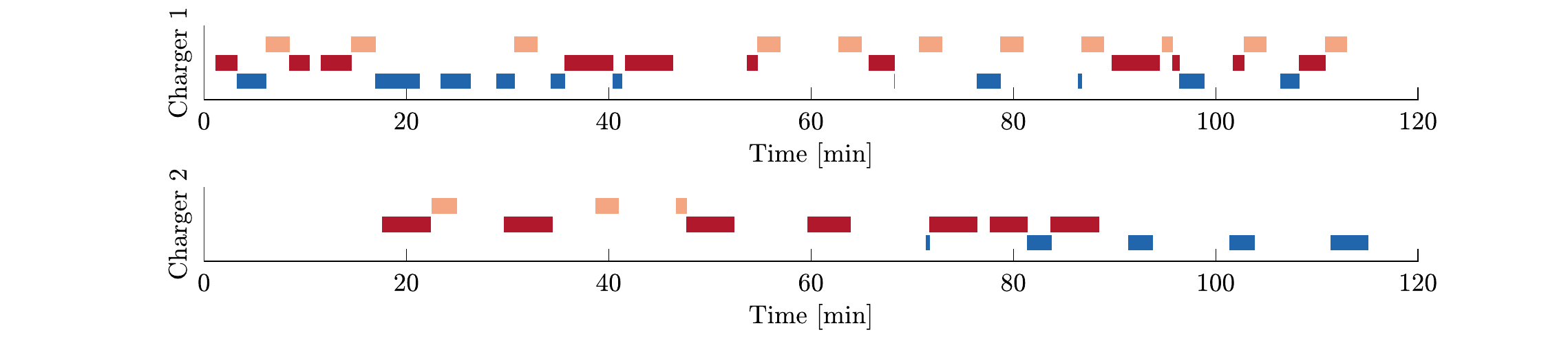} \label{fig:heuristic1}}%
\vfil
\subfloat[Improved heuristic.]
{\includegraphics[width=\textwidth]{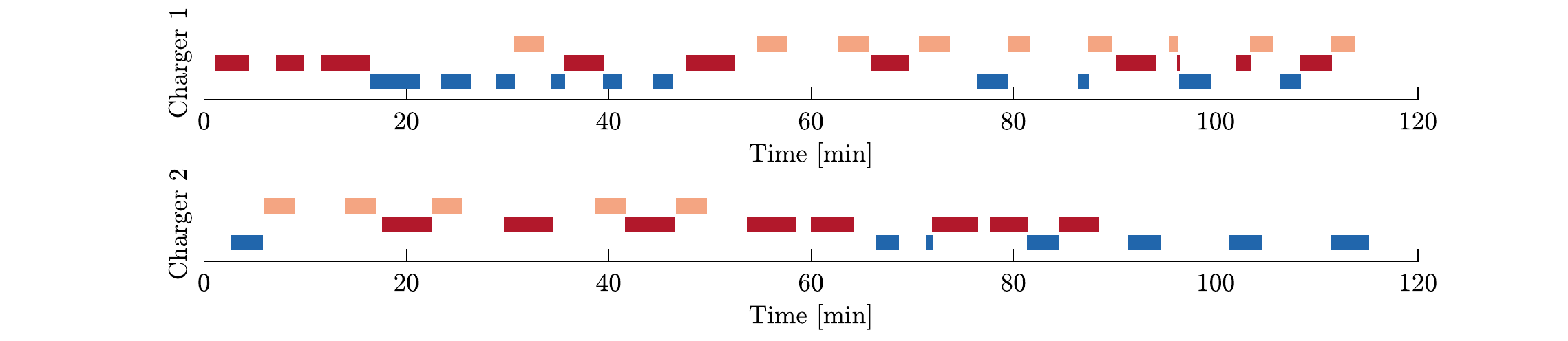} \label{fig:heuristic2}}%
\caption{Example of planned charger use timelines in a scenario with 2 chargers, 3 bus lines, and an horizon length of 2 hours. Each bus line is denoted by a different color, and each rectangle in the charger timelines represents one planned charging event.}
\label{fig:heuristics}
\end{figure}

\subsection{Local search heuristic} \label{sec:4.3}

Let us assume that a solution $z(\beta^\kappa)$ to the optimization problem \eqref{eq:dual_function} has been computed by solving all the bus line subproblems. We now propose an intuitive and computationally inexpensive local search heuristic to construct a feasible solution of \eqref{eq:MILP} $z_{\textup{feas}}^\kappa$. In essence, this means that $z(\beta^\kappa)$ has to be modified enough to satisfy the relaxed constraints \eqref{eq:exclusion1}-\eqref{eq:exclusion2}. This amounts to solving any charging conflicts that may occur in $z(\beta^\kappa)$, that is, make sure that no planned charging events are ever overlapping.

The core idea of the proposed heuristic is to keep the same charging schedule as in $z(\beta^\kappa)$ by considering a version of MILP \eqref{eq:MILP} where all the values of the binary variables have been fixed. The charging decision variables in the vector $b$ can simply be fixed to their values in $z(\beta^\kappa)$. Then, one way to fix the values of the charging order variables in $\psi$ from there is, for any pair of terminal visits ${\{l,i,k\}}$ and ${\{l',i',k'\}}$, to compare the times $t^{\text{char},k}_{l,i}$ and $t^{\text{char},k'}_{l',i'}$ at which charging is planned to begin in $z(\beta^\kappa)$, and fix the value of $\psi_{\{l,i,k\},\{l',i',k'\}}$ accordingly. Once all binary variables have been fixed, the problem obtained has the same objective function and constraints as \eqref{eq:MILP}, but without the binary constraints in \eqref{eq:binary_in_MILP} and \eqref{eq:exclusion_in_MILP}. The problem obtained is therefore an LP, and can be solved quite fast with a dedicated solver.

This basic heuristic can provide feasible solutions, but it is easy to see how the quality of these solutions can be further improved by carrying out small modifications of the charging schedule in $z(\beta^\kappa)$. Indeed, in case several chargers are available to chose from, it makes no difference which exact charger is used in the absence of charger exclusion constraints when solving problem \eqref{eq:dual_function}. As a result, the charging schedule in $z(\beta^\kappa)$ might include periods during which one charger is unnecessarily loaded while the other ones are all available. An example of this is given in Figure \ref{fig:heuristic_infeasible}, where most charging events have been arbitrarily assigned to the first charger. Since the charging decision variables are directly fixed to their values in $z(\beta^\kappa)$ with the basic heuristic, this arbitrary choice of chargers is embedded in the resulting feasible solution. This can be seen in Figure \ref{fig:heuristic1}, where all charging event overlaps have indeed been eliminated, but where the first charger is still being overused.

The proposed basic heuristic can then be improved in the following way. First, all charging event overlaps in $z(\beta^\kappa)$ are identified and sorted in order of decreasing overlap duration. Then, for each overlap in the charging schedule, we check if any of the two charging events involved could be moved to another charger without causing any additional overlap there. If so, then the conflict is resolved by transferring that charging event to another charger, and the charging decisions variables in the vector $b$ are modified accordingly. This way, the charging load is distributed more evenly between chargers, which improves the quality of the starting point from which feasible solutions are constructed. Once this initialization phase is completed, the same process as above is carried out: all binary variables are fixed and an LP is solved to optimality to provide a feasible solution. Figure \ref{fig:heuristic2} shows the planned charger use timelines for a feasible solution constructed with this improved heuristic. It can be observed that charging event overlaps have been successfully removed, and that some of the charging events have now been transferred to the second charger, as opposed to the solution of the basic heuristic in Figure \ref{fig:heuristic1}. The charging schedule is therefore more balanced with the improved heuristic, which results in planned charging events which are on average longer as can be seen in the figures. Consequently, the improved heuristic is the one we use to generate feasible solutions in the Lagrangian heuristic procedure.

\section{Simulations} \label{sec:5}

In this section, two distinct simulation studies are presented. First, a scalability study is run on synthetic bus networks of different sizes in order to assess how well the Lagrangian heuristic in the proposed integrated control architecture scales compared with similar methods. Then, the actual performance of the proposed integrated control architecture is evaluated in a case study based on a detailed model of three bus routes from the Chicago transit network in a microscopic simulation environment.

\subsection{Scalability analysis}

Synthetic simulations are carried out to evaluate the computational performance of the proposed Lagrangian heuristic. In particular, the Lagrangian heuristic is compared to an off-the-shelf solver deployed directly on the original MILP \eqref{eq:MILP}. The well-established solver Gurobi has been used in this work. By default, Gurobi uses the branch-and-cut method to solve MILPs to optimality. In case the search procedure is interrupted prematurely, when a maximum runtime limit is hit for instance, Gurobi returns the best known incumbent together with the current optimality gap. This enables comparisons with the best feasible solution and optimality gap returned by the Lagrangian heuristic. It should also be mentioned that Gurobi is used in the Lagrangian heuristic procedure to solve both the bus line subproblems \eqref{eq:subproblem} and the LPs appearing in the local search heuristic. Contrary to the original MILP \eqref{eq:MILP}, the subproblems are much smaller MILPs that can be solved to optimality in a reasonable time, usually a few seconds, with Gurobi. Since the subproblems are independent, they can in principle be solved in parallel. Therefore, only the largest computation time needed when solving each subproblem is considered at each iteration when evaluating the runtime of the Lagrangian heuristic.

On top of the Lagrangian heuristic and the direct use of Gurobi, we also investigate the performance of a heuristic based on the linear programming (LP) relaxation of MILP \eqref{eq:MILP}, which is referred to as \textit{LP heuristic} hereafter. The LP relaxation of \eqref{eq:MILP} is formed by letting all binary variables in this problem take values in the continuous interval $[0,1]$ instead. The solution to this LP is infeasible in the original MILP in general, but does provide a lower bound on the optimal value of \eqref{eq:MILP} and can be used as the starting point of a local search heuristic to build a feasible solution, in the same spirit as in the Lagrangian heuristic. The LP heuristic thus consists of two stages: the LP relaxation of \eqref{eq:MILP} is first solved to optimality with Gurobi, and a feasible solution is then built using the heuristic presented in Section \ref{sec:4.3}.

Simple synthetic bus networks are constructed for this scalability study. Bus routes of different lengths, with different numbers of bus stops, and different target headways are considered. Bus stops are distributed uniformly along each bus route, with the initial stop being the bus terminal shared by all bus lines to comply with our modeling assumptions. All chargers are assumed to be located directly at the bus terminal. Passenger arrival rates and alighting proportions are assumed to be the same at all bus stops. Travel time bounds are derived directly from the speed limit and minimum speed allowed, which are set to 50 km/h and 30 km/h respectively. In order for these synthetic bus networks to be realistic, they are built by using the same basic simulations parameters as for the Chicago case study. We refer the reader to Table \ref{tb:parameters}, found in the next subsection, for reference.

\begin{table*}[!t]
\renewcommand{\arraystretch}{1.1}
\caption{Computational performance of the methods tested for transit networks of increasing size. \textit{Lagrange} denotes the Lagrangian heuristic while \textit{Gurobi} denotes the direct use of Gurobi on MILP \eqref{eq:MILP}.}
\label{tb:scalability}
\centering
{\begin{tabular}{|ll|c|c|c|c|} \hline
Network & & $\#$ buses & $\#$ stops & $\#$ variables & $\#$ constraints \\
$n_L=8$ & $n_c=6$ & $53$ & $187$ & $31\,000$ & $79\,000$ \\ \hline
& & Best incumbent & Lower bound & Optimality gap & Runtime \\ \hline
Lagrange & $n_{\text{iter}}=2$ & $500$ & $291$ & $40.9\%$ & $29$ s \\ 
& $n_{\text{iter}}=3$ & $473$ & $366$ & $22.5\%$ & $43$ s \\ 
& $n_{\text{iter}}=5$ & $469$ & $384$ & $18.0\%$ & $66$ s \\ \hline
Gurobi & $t_{\text{lim}}=300$ s & $1001$ & $382$ & $60.0\%$ & $300$ s \\ 
& $t_{\text{lim}}=1800$ s & $468$ & $382$ & $18.3\%$ & $1800$ s \\ \hline
LP heuristic & & $486$ & $378$ & $22.2\%$ & $19$ s \\ \hline \hline
Network & & $\#$ buses & $\#$ stops & $\#$ variables & $\#$ constraints \\
$n_L=12$ & $n_c=9$ & $84$ & $276$ & $50\,000$ & $200\,000$ \\ \hline
& & Best incumbent & Lower bound & Optimality gap & Runtime \\ \hline
Lagrange & $n_{\text{iter}}=2$ & $660$ & $313$ & $52.6\%$ & $38$ s \\ 
& $n_{\text{iter}}=3$ & $660$ & $522$ & $20.9\%$ & $55$ s \\ 
& $n_{\text{iter}}=5$ & $653$ & $556$ & $14.9\%$ & $89$ s \\ \hline
Gurobi & $t_{\text{lim}}=300$ s & $1648$ & $553$ & $66.4\%$ & $300$ s \\ 
& $t_{\text{lim}}=1800$ s & $1287$ & $553$ & $56.7\%$ & $1800$ s \\ \hline
LP heuristic & & $1336$ & $546$ & $59.0\%$ & $32$ s \\ \hline \hline
Network & & $\#$ buses & $\#$ stops & $\#$ variables & $\#$ constraints \\
$n_L=15$ & $n_c=11$ & $104$ & $334$ & $66\,000$ & $354\,000$ \\ \hline
& & Best incumbent & Lower bound & Optimality gap & Runtime \\ \hline
Lagrange & $n_{\text{iter}}=2$ & $834$ & $240$ & $71.2\%$ & $41.5$ s \\ 
& $n_{\text{iter}}=3$ & $829$ & $629$ & $24.1\%$ & $65$ s \\ 
& $n_{\text{iter}}=5$ & $809$ & $684$ & $15.5\%$ & $109$ s \\ \hline
Gurobi & $t_{\text{lim}}=300$ s & $2102$ & $680$ & $67.6\%$ & $300$ s \\ 
& $t_{\text{lim}}=1800$ s & $1884$ & $681$ & $63.7\%$ & $1800$ s \\ \hline
LP heuristic & & $1117$ & $673$ & $39.2\%$ & $43$ s \\ \hline \hline
Network & & $\#$ buses & $\#$ stops & $\#$ variables & $\#$ constraints \\
$n_L=20$ & $n_c=14$ & $132$ & $444$ & $95\,000$ & $753\,000$ \\ \hline
& & Best incumbent & Lower bound & Optimality gap & Runtime \\ \hline
Lagrange & $n_{\text{iter}}=2$ & $1100$ & $206$ & $81.0\%$ & $46$ s \\ 
& $n_{\text{iter}}=3$ & $1094$ & $763$ & $30.6\%$ & $81$ s \\ 
& $n_{\text{iter}}=5$ & $1071$ & $887$ & $17.4\%$ & $139$ s \\ \hline
Gurobi & $t_{\text{lim}}=300$ s & $2759$ & $882$ & $68.1\%$ & $300$ s \\ 
& $t_{\text{lim}}=1800$ s & $2734$ & $882$ & $67.8\%$ & $1800$ s \\ \hline
LP heuristic & & $1501$ & $872$ & $39.9\%$ & $91$ s \\ \hline
\end{tabular}}
\end{table*}

A simple version of the problem \eqref{eq:MILP} is also considered. Buses are initialized with full batteries, no passengers onboard, and outside rush hours. The goal of this study is to get an estimate of how well each approach scales on larger transit systems rather than to establish precisely the runtime on each problem. Hence, it is enough to consider these simple versions of the optimization problem \eqref{eq:MILP} since they capture its main features, even if they might represent an idealized state of a real transit system.

For every bus network, several versions of the problem are formed by choosing random starting positions for the vehicles each time. These different versions of the problem are the same for every method studied. Table \ref{tb:scalability} presents the main results of this study averaged over 10 repetitions for each bus network, with bus networks consisting of 6 to 20 bus lines. In this table, $n_{\text{iter}}$ denotes the number of subgradient iterations and $t_{\text{lim}}$ the maximum runtime allowed when trying to solve MILP \eqref{eq:MILP} with Gurobi directly. Results for only up to 5 iterations of the subgradient algorithm are shown for the Lagrangian heuristic as no significant improvement in performance has been observed past that point, due to the generally fast initial convergence of the subgradient algorithm.

Very good lower bounds can be obtained when trying to solve MILP \eqref{eq:MILP} with Gurobi directly, almost as good as the ones found by the Lagrangian heuristic, as can be seen in the third column in Table \ref{tb:scalability}. On the other hand, Gurobi struggles to find good quality solutions, judging from the second column. This is especially true for the larger networks, where the best incumbents found by Gurobi are poor, even for long computation times of up to 30 minutes. The lower bound from the LP relaxation is usually pretty good on these problems and does not seem to be too far from the Lagrangian relaxation bound, as can be seen in the third column of the table. However, the feasible solutions built from the solutions of the LP relaxation cannot compete with the feasible solutions returned by the Lagrangian heuristic, even when it is run with as few as 2 iterations. In fact, it can be observed from Table \ref{tb:scalability} that the Lagrangian heuristic closes the gap from below rather than from above like Gurobi is doing: the Lagrangian heuristic is very good at generating good quality solutions very early on but getting a good lower bound usually requires a few iterations. For these problems, at least 3 iterations of the Lagrangian heuristic seem to be needed in order to obtain good lower bounds.

It can be observed from the fourth and fifth columns in Table \ref{tb:scalability} that the growth in the number of variables and constraints is superlinear, because the number of potential pairs of terminal visits increases exponentially with the number of bus lines. This in turn results in a superlinear increase in the number of constraints and variables in the problem \eqref{eq:MILP} due to the charger exclusion constraints \eqref{eq:exclusion1}-\eqref{eq:exclusion2} and the associated charging order variables. The computation times for each method are presented in the last column of Table \ref{tb:scalability}. The runtime of the LP heuristic corresponds to roughly between 1 and 3 iterations of the Lagrangian heuristic, depending on the problem size, where the LP heuristic gets slower relative to the Lagrangian heuristic on larger problem instances. Even though the LP heuristic can be used to generate a good lower bound quite fast, the second column of the table suggests that the associated feasible solution is much worse than the one that can be obtained from the Lagrangian heuristic in a similar time, with the exception of the smallest problem considered. Finally, Gurobi cannot be considered a practically viable alternative due to its very long computation times across all the networks studied.

Therefore, the Lagrangian heuristic seems to be the most adapted method for generating good feasible solutions in short amounts of time, while the LP heuristic appears to be the most adapted method for quickly finding good lower bounds. In case longer runtimes are acceptable, more iterations of the subgradient algorithm can be performed and the Lagrangian heuristic outperforms the LP heuristic on all the metrics considered. It must be noted, however, that large problems where 20 bus lines or more are considered simultaneously require around one to a few minutes runtime for the Lagrangian heuristic to reach good optimality gaps. This may or may not be acceptable, depending on the frequency at which the high-level references are planned to be updated in the control framework, but it is nonetheless interesting to note that this is where the proposed Lagrangian heuristic might begin to show its limitations.

\subsection{Case study}

We now move on to the main case study of this work. We consider a subset of the transit network of the city of Chicago consisting of three bus routes, namely routes 7, 12, and 85. These bus lines are operated by the Chicago Transit Authority, the main transit provider in Chicago, who has compiled a publicly-available dataset \cite{CTA} with ridership information for several of the bus lines it operates during the month of October 2012. This passenger data was used to calibrate the arrival rates and the alighting proportions of passengers in our model. Each of these three bus lines theoretically services over one hundred bus stops, since most road intersections can be used as bus stops. However, many such stops have very few if any passengers boarding or alighting, and so it was decided to restrict the set of bus stops to the ones with at least 6 boarding or alighting passengers per hour, i.e. with a new passenger passing through every 10 minutes on average. The stops that failed to meet this simple criterion were discarded from the route models. The stop of Harrison $\&$ Central, which is an end stop for all of these three routes and also their only point of overlap, was assumed to be the shared terminal at which chargers are located. Note, however, that none of these bus routes were electrified at the time of data collection, and that no opportunity charging infrastructure was located at Harrison $\&$ Central. We believe that it is nonetheless valuable to base our case study on these bus routes, partly because the main features of the bus lines modeled would be the same regardless of the type of buses deployed, and partly because this study can then function as a test bed to investigate the operational implications of electrifying an existing bus network. These three bus routes have been chosen in part because they go through different driving environments: bus routes 7 and 12 lie along an east-west axis between the dense city center of downtown Chicago and the suburban area where the terminal is located, while bus route 85 lies along a north-south axis in a fully suburban environment. Consequently, buses of line 85 encounter less passengers and traffic than buses from the other lines, and their target headway is also higher.

\begin{figure}[b]
\begin{center}
\includegraphics[width=\textwidth]{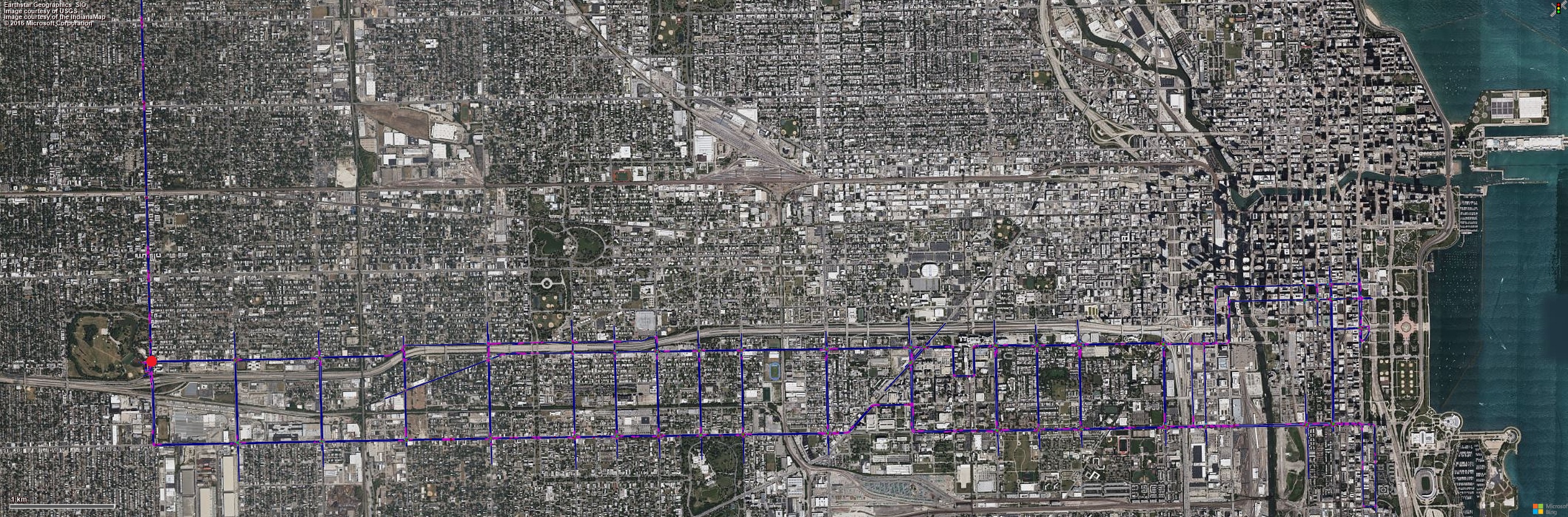}
\caption{Layout of bus routes 7, 12, and 85 in Vissim shown on an aerial view of Chicago. The red circle on the left symbolizes the shared bus terminal. Bus routes 7 and 12, which are parallel to each other, stretch all the way to downtown Chicago on the right. Only a fraction of route 85 is represented above the bus terminal.}
\label{fig:vissim}
\end{center}
\end{figure}

Bus routes 7, 12, and 85 have been modeled in the microscopic traffic simulator Vissim, as is shown in Figure \ref{fig:vissim}. The passenger flows at bus stops were calibrated from the dataset mentioned above. In addition, the traffic flows on the roads trafficked by the buses were calibrated from historical traffic flow measurements available through the website of the city of Chicago \citep{Chicago}. This data was used to set the input vehicle rates for each of the traffic links modeled in the simulation environment. No information on the traffic light cycles could be obtained, however, and simple 30-second cycles were chosen instead.

For each inter-stop link modeled in the Vissim simulation environment, samples are taken for different travel time commands and bus mass values in order to estimate the energy consumption function. The energy consumption corresponding to each sample is calculated by using the same longitudinal dynamics, electric bus powertrain, and battery models as in \cite{lacombe2}. Then, the piecewise linear approximation \eqref{eq:energy_consumption} with travel time breakpoints spaced evenly is fitted to the result. Since Chicago is a very flat city, the road gradient used in the longitudinal dynamics model was assumed to be zero for all inter-stop links. As a result, the energy consumption profiles are quite simple, and a 2-piece approximation was found to be sufficient for all links. Similarly, the travel time bounds $T^{\text{min},k}_{l,j}$ and $T^{\text{max},k}_{l,j}$ appearing in \eqref{eq:time_bounds} are calibrated as the mean travel time values of samples where the speed reference for buses is set to the minimum acceptable speed of 30 km/h, or the speed limit of 50 km/h, which applies on all the links, respectively.

\begin{figure}[b]
\begin{center}
\includegraphics[width=\textwidth]{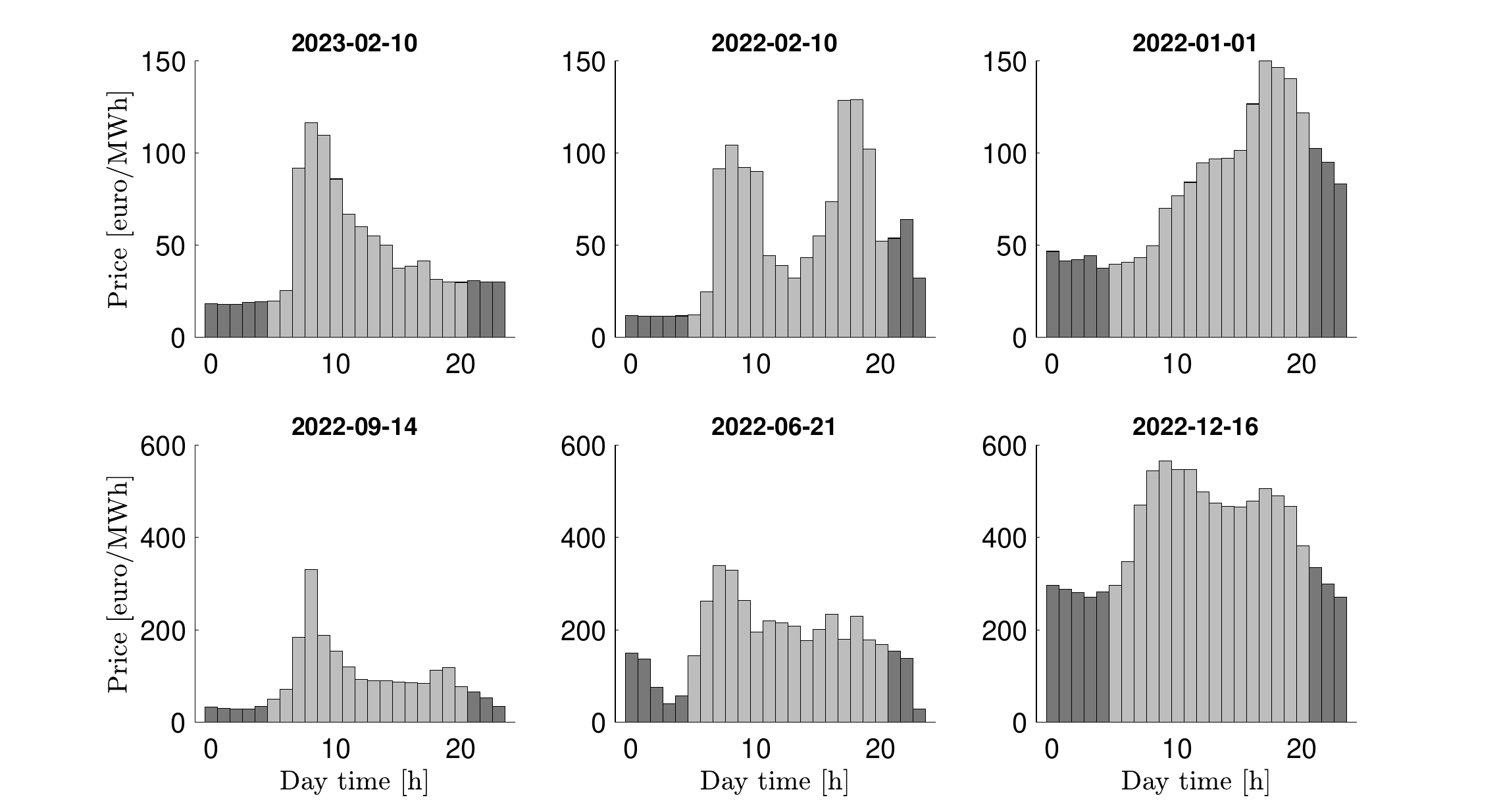}
\caption{Time-of-use electricity pricing profiles used in the case study. Each bar represents the electricity price for one hour of the day. The profiles are sorted in order of increasing daily average electricity prices. The electricity prices during the hours of the night not included in the simulations are shown in dark gray. Only the light gray bars are relevant in this case study.}
\label{fig:tou_profiles}
\end{center}
\end{figure}

Historical time-of-use pricing data were obtained from the European power exchange platform Nordpool. Since case study simulations are run for full days of bus operations, a sample of daily electricity price profiles has been selected from Nordpool. Note that time-of-use price profiles are always available one day ahead in the European market. The 6 profiles selected in this case study are shown in Figure \ref{fig:tou_profiles}. We have tried to pick distinct profiles that cover the widest range of possible price variations. Some profiles have a very low average price while others have a very high average price, some are relatively constant while others have sharp variations during the day, etc.

\begin{table*}[!h]
\renewcommand{\arraystretch}{1.1}
\caption{Simulation parameters.}
\label{tb:parameters}
\centering
{\begin{tabular}{|l|l|l|l|l|l|} \hline
$T=2$ h & $n_L=3$ & $n_1^b=11$ & $\sigma_\text{ini}=1$ & $P_{\text{char}}=300$ kW & $Q=264$ kWh \\
$T_{\text{sim}}=16$ h & $n_c=2$ & $n_2^b=14$ & $\sigma_\text{end}=0.3$ & $m_\text{cap}=19500$ kg & $m_\text{pax}=60$ kg \\
$N=16$ & $n_q=2$ & $n_3^b=7$ & $\sigma_\text{min}=0.3$ & $d_\text{char}=10$ s & $d_\text{pax}=1.5$ s \\
$M=10^5$ & $H_1=5$ mn & $n_1^s=23$ & $p_\text{reg}=0.0047$ \euro/s & $c_{\text{fixed},1}=231$ s & $E_0=0$ kWh \\
$n_\text{iter}=5$ & $H_2=6$ mn & $n_2^s=67$ & $p_\text{end}=0.4$ \euro & $c_{\text{fixed},2}=175$ s & \\
$\epsilon=2$ & $H_3=8$ mn & $n_3^s=44$ & $p_\text{reject}=100$ \euro & $c_{\text{fixed},3}=223$ s & \\ \hline
\end{tabular}}
\end{table*}

All the remaining numerical values of the simulation parameters used in this case study can be found in Table~\ref{tb:parameters}. Note that the battery capacity is assumed to be the same for all vehicles here and noted $Q$. Similarly, $\sigma_{\text{min}}$ is the minimum state-of-charge value allowed for all bus lines and is set to $30\%$. Not only does this leave a comfortable margin for buses to be able to complete one full trip back to the terminal, it also prevents batteries from being emptied too much, thus improving their life expectancy by avoiding deep discharges. The 12 m version of the Volvo 7900 electric bus \citep{Volvo} is used as a reference to set the battery capacity value $Q=264$ kWh, the opportunity charging power $P_{\text{char}}=300$ kW and the bus capacity $m_\text{cap}=19500$ kg. The horizon length $T$ is set to 2 hours as this was found to be a good trade-off between long prediction horizons and reasonable computation times. The number of buses deployed on each route has been calibrated experimentally, based on the desired target headways, to be the minimum number of buses required to be able to maintain the desired headways while still allowing some time for charging at the terminal. The number of available chargers at the terminal has then been set to $n_c=2$ as it has been observed experimentally that a single charger was not enough to cover the resulting charging demand, and that using 3 chargers resulted in a lot of charger idle time and could therefore not be justified. The pricing parameter for headways deviations $p_{\text{reg}}$ is based on the passenger value of time, i.e. what monetary value passengers would give to the time they spend waiting. Following \cite{petit}, we choose to set the passenger value of time to about half of the average hourly income in Chicago. The pricing parameter for the terminal cost $p_{\text{end}}$ is chosen to be several times greater than the mean electricity price to encourage sufficient use of the chargers. Finally, the pricing parameter for refusing passenger boarding $p_{\text{reject}}$ is set to a very high value so as to keep it as a last resort measure to be used only to satisfy bus capacity constraints.

\begin{figure}[b]
\begin{center}
\includegraphics[width=0.8\textwidth]{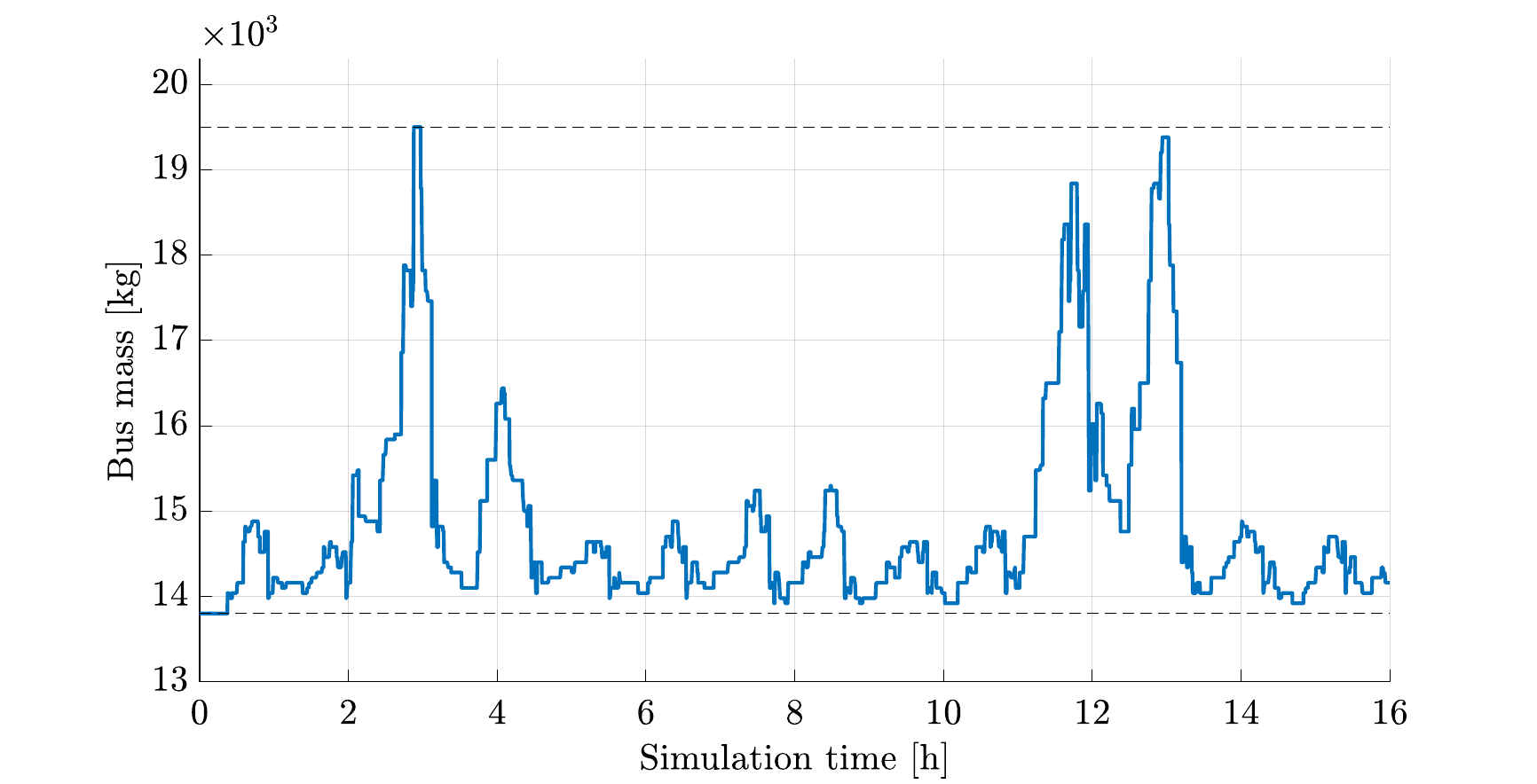}
\caption{Example of the mass trajectory of one bus in one of the simulations. The dashed lines represent the mass of the bus when it is empty and when it is full.}
\label{fig:mass}
\end{center}
\end{figure}

One day of bus operations is assumed to last for 16 hours in the simulations, the 8 remaining hours being devoted to overnight charging for most of the vehicles. It is assumed that each simulation begins at 5 AM and ends at 9 PM. In order to represent public transit operations faithfully, one morning and one evening rush hours are considered. The morning rush hour is assumed to take place between 7 AM and 9 AM, and the evening rush hour between 4 PM and 6 PM. During rush hours, the passenger arrival rates are multiplied by 4 and the vehicle input rates are multiplied by 2 in the simulation environment in order to account for the increased number of road users and passengers. Figure \ref{fig:mass} gives an idea of how these increased passenger loads can affect bus operations. This figure displays the evolution of the mass of one bus of line 7 in one of the simulations. The increased passenger loads during rush hours can be clearly seen in this figure, and the bus capacity is even reached at one point. Note that rush hours take place between the 2nd and 4th hours, and the 11th and the 13th hours of simulation time.

In addition to the proposed integrated control architecture, the performance of two rule-based control baselines are studied. These baseline methods do not rely on predictive control nor on solving optimization problems and aim at simulating the current best practices currently implemented in the field by transit agencies. The operational control and charging scheduling aspects are handled in the following way by these baseline methods:
\begin{itemize}
    \item In both methods, operational decisions are taken based on the information currently available so as to be \textit{locally optimal}. By this, we mean that the travel time decisions when leaving each bus stop are taken so as to reach the next stop exactly one target headway after the preceding bus has reached it. Only the information on the arrival times of the preceding bus is needed to implement this type of control. Similarly, holding times are chosen so as to dispatch buses from the terminal at one target headway from one another whenever possible.
    \item Charging decisions are taken on a first-come, first-served basis in the two baseline methods: buses use whichever charger is available next and can accommodate their desired charging duration. The two baselines differ in how the charging times are chosen. In the first baseline, referred to as the \textit{adaptive baseline} hereafter, buses are made to charge long enough to reach the current desired state-of-charge value $\sigma_\text{goal}$, as outlined in Section \ref{sec:3.3}. This way, the adaptive baseline can be compared on an equal footing with the proposed control architecture, since they pursue the same charging goals. Note that in this case, the value of $\sigma_\text{goal}$ used in the baseline must be shifted by the length of the control horizon used in the proposed control method $T$, since $\sigma_\text{goal}$ was initially designed to be the desired value at the end of the horizon. The second baseline, called the \textit{static baseline}, is meant to be closer to current transit agency practices by relying on fixed charging times. Each bus line is assumed to have its own fixed charging time, which is set based on the daily average energy consumption on that line and on the frequency of terminal visits. The fixed charging times are then designed with the goal of having buses reach the minimum state-of-charge value allowed $\sigma_\text{min}$ at the end of the day when starting the day with full batteries. Each time a bus reaches the terminal, it is made to charge for the fixed charging time of the bus line to which it belongs. For this case study, the fixed charging times for all bus lines can be found in Table \ref{tb:parameters}, where they are noted as $c_\text{fixed}$.
\end{itemize}
Note that, once travel time, holding, and charging decisions have been taken by the baselines, these decisions are tracked by a lower-level control layer in exactly the same way as for the proposed control architecture.

\begin{figure}[b]
\begin{center}
\includegraphics[width=\textwidth]{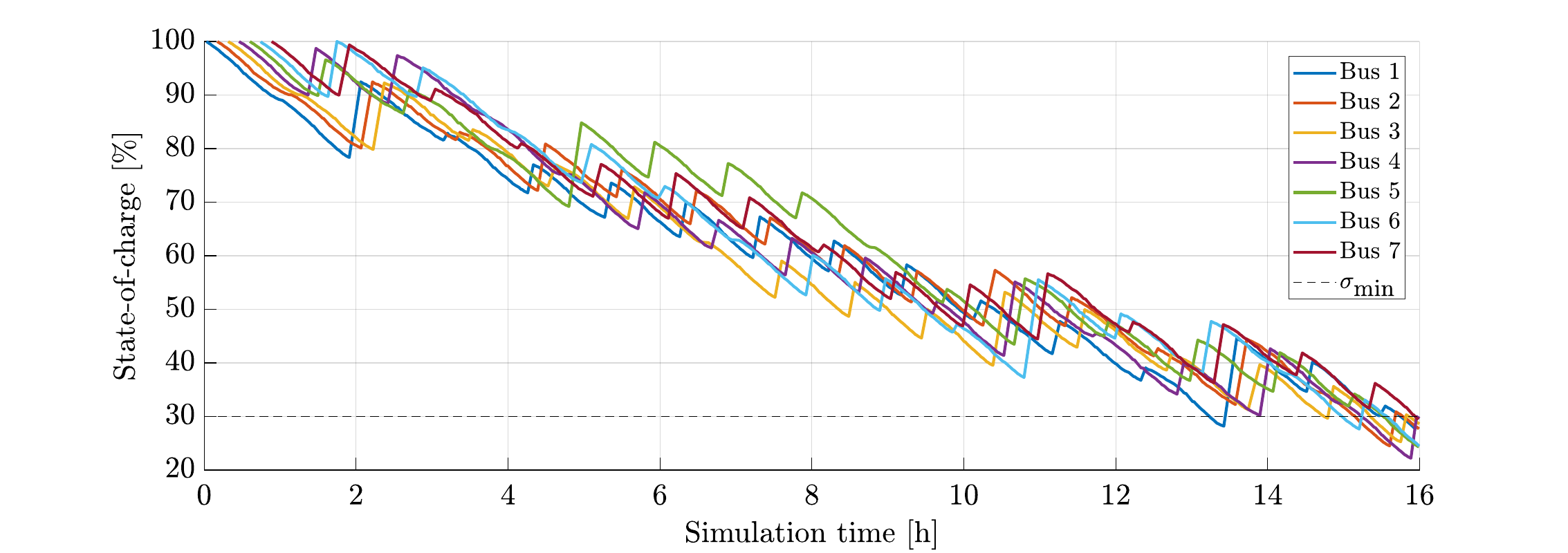}
\caption{Example of the state-of-charge trajectories for buses of line 85 in one of the simulations using the 2022-06-21 time-of-use pricing profile and with the integrated control architecture. The dashed line represents the minimum state-of-charge value allowed $\sigma_\text{min}$.}
\label{fig:SoC}
\end{center}
\end{figure}

\begin{figure}[b]
\begin{center}
\includegraphics[width=\textwidth]{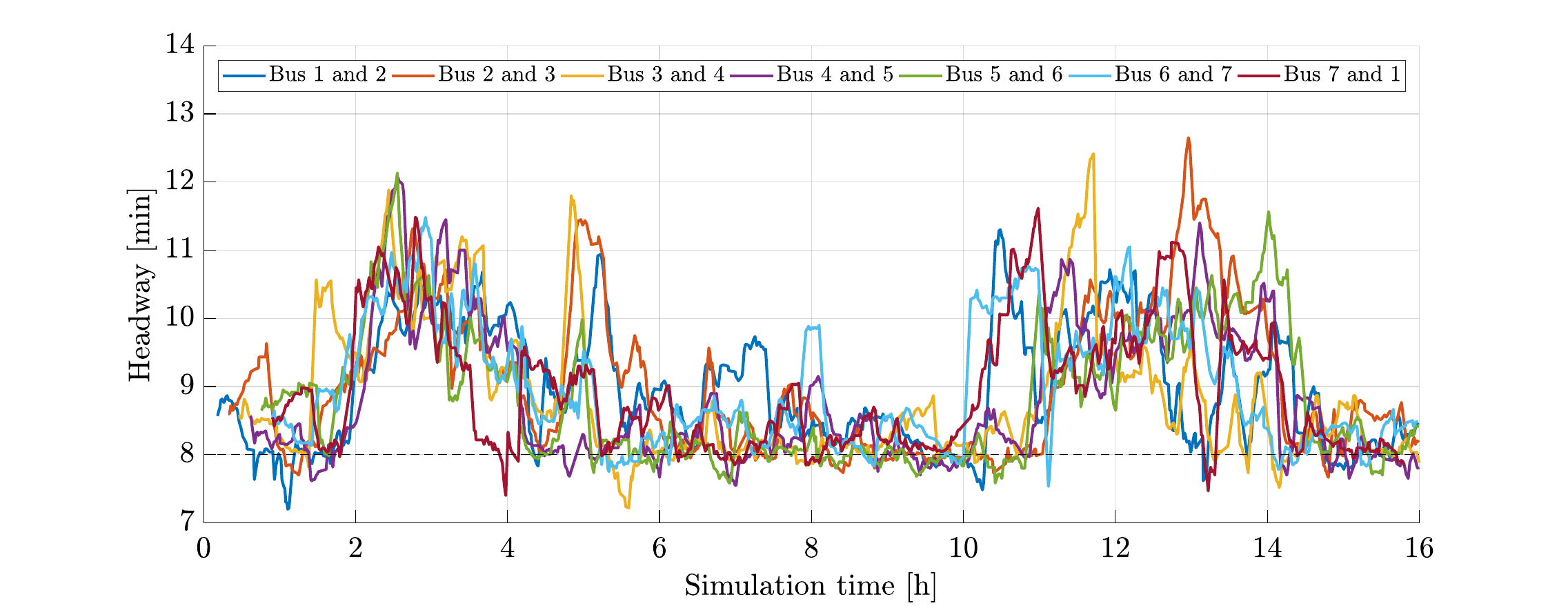}
\caption{Example of bus headways for line 85 in one of the simulations using the 2022-06-21 time-of-use pricing profile and with the integrated control architecture. The target headway, which is 8 minutes for bus line 85, is materialized by the dashed line.}
\label{fig:headways}
\end{center}
\end{figure}

\begin{figure}[b]
\centering
\subfloat[Static baseline.]
{\includegraphics[width=\textwidth]{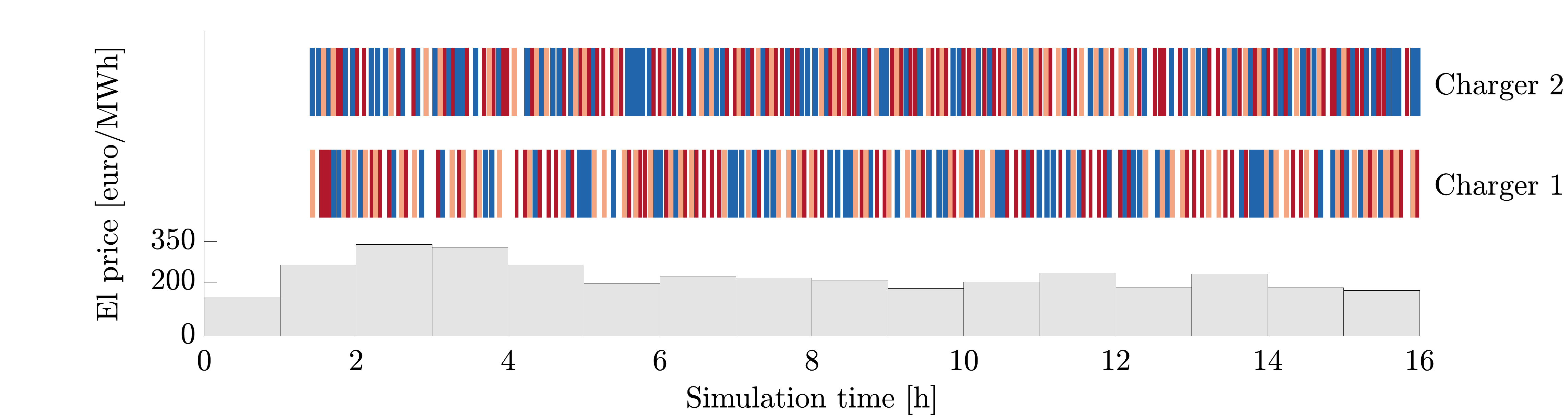} \label{fig:charger_timeline_baseline}}%
\vfil
\subfloat[Integrated control.]
{\includegraphics[width=\textwidth]{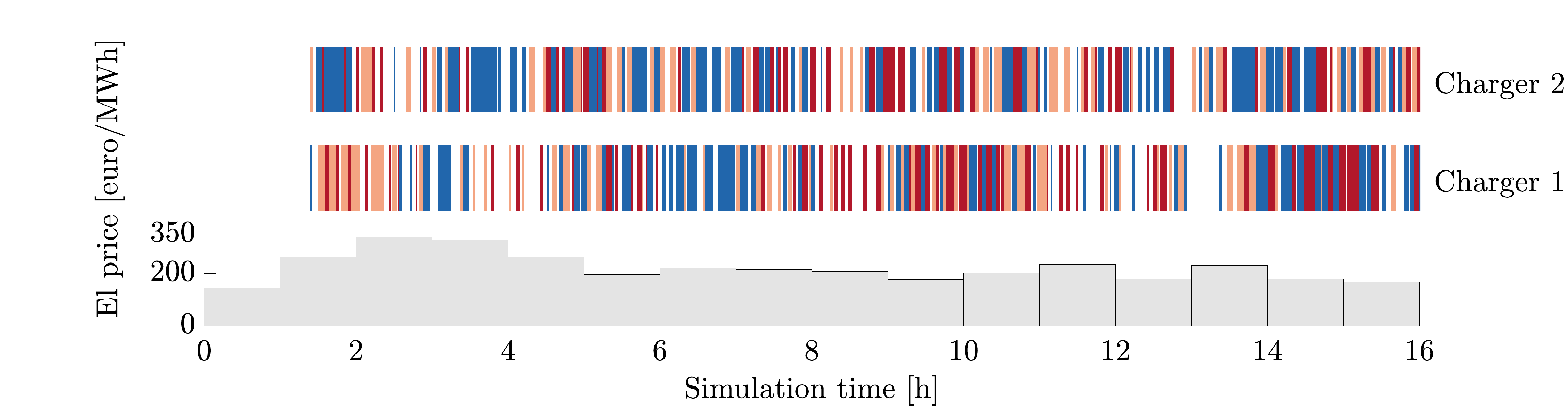} \label{fig:charger_timeline_controller}}%
\caption{Charger use timelines for one full simulation using the 2022-06-21 time-of-use pricing profile. Each bus line is denoted by a different color: blue for line 7, red for line 12, and pink for line 85. Each rectangle in the charger timelines represents one charging event for a bus of the line of the corresponding color. The electricity prices are plotted for reference.}
\label{fig:charger_timeline}
\end{figure}

\begin{figure}[b]
\begin{center}
\includegraphics[width=\textwidth]{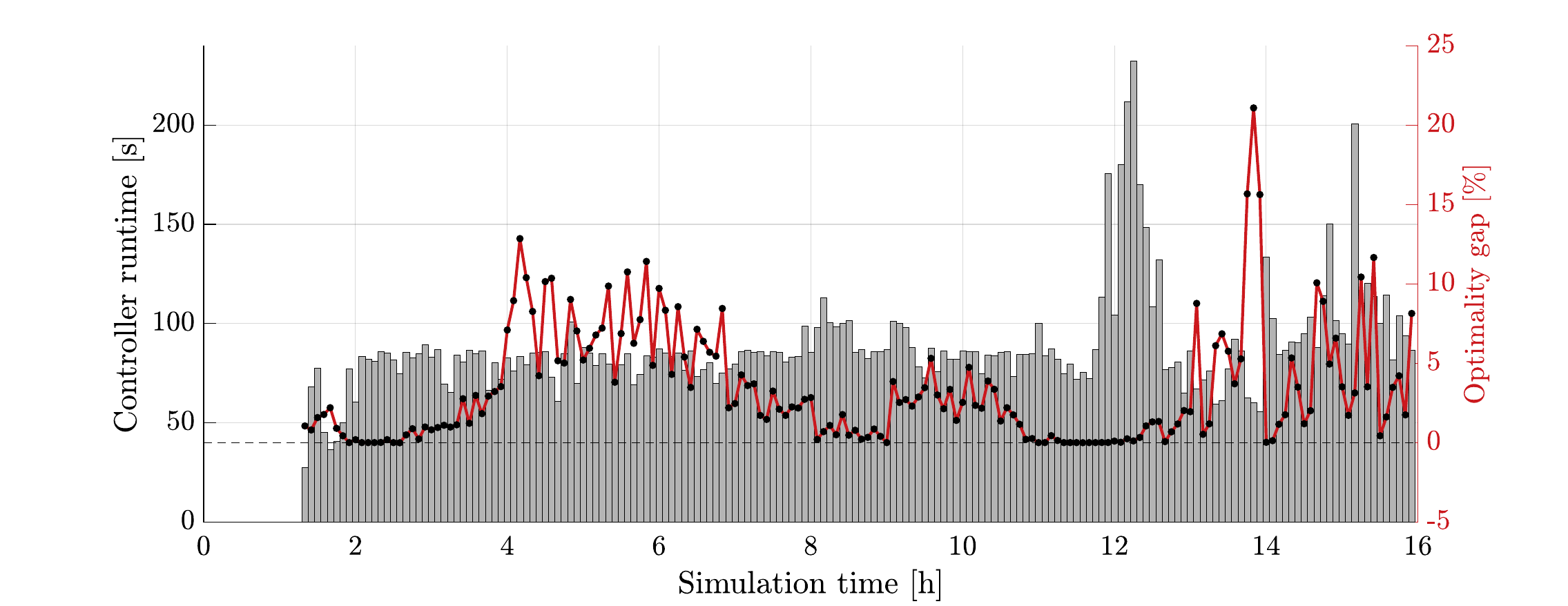}
\caption{Overview of the computational performance of the Lagrangian heuristic for one of the simulations. The gray bars on the left-hand side y-axis indicate the total runtime needed for solving each problem instance. The corresponding optimality gaps achieved by the Lagrangian heuristic on each problem instance are plotted as black circles on the left-hand side y-axis. The dashed line indicates a zero optimality gap.}
\label{fig:optimization}
\end{center}
\end{figure}

\subsection{Results}

Full-day simulations are carried out for each of the time-of-use charging profiles presented in Figure \ref{fig:tou_profiles}, with 5 repetitions for each of them. In each simulation, the transit system is left to run on its own for 1 hour and 20 minutes of simulation time with only minimum control before the controllers studied start being applied. This means, among other things, that no charging event takes place before that time, and that costs are only counted from that point on. Then, either the integrated control architecture or one of the baseline controllers is deployed. High-level references are updated at a frequency of 5 minutes in the integrated control architecture. State-of-charge values which are above $\sigma_\text{min}$ at the end of the simulation are discounted from the charging cost at half the average daily electricity price.

First, results from the same simulation instance using the 2022-06-21 time-of-use pricing profile are shown in Figures \ref{fig:SoC}, \ref{fig:headways}, \ref{fig:charger_timeline}, and \ref{fig:optimization} in order to illustrate some key observations. Figure \ref{fig:SoC} shows the state-of-charge trajectories and Figure \ref{fig:headways} the headway trajectories obtained with the proposed integrated control architecture on this simulation instance. Note that all the trajectories displayed in these two figures begin at different times since buses are introduced in the simulation one by one initially. The influence of rush hours on the charging decisions can be seen in Figure \ref{fig:SoC}, where very little charging is carried out between the 3rd and 4th simulation hours, and between the 12th and 13th simulation hours, since buses need to be picking up the additional passengers and to be catching up on potential delays incurred by the increased amount of traffic. Observe also that some of the state-of-charge values go below $\sigma_\text{min}$ towards the end of the simulation in this figure. This is something that vehicles are allowed to do since constraint \eqref{eq:SoC_feasibility} imposes that buses have a state-of-charge greater than $\sigma_\text{min}$ only when leaving the terminal. Figure \ref{fig:headways} shows the headway trajectories over a day of traffic operation together with the target headway that the controller is trying to reach. It can be seen that the deviations from the target headway are largest during rush hours and one to two hours after since it takes some time for regular service to be restored, but that the controller manages to keep fairly small deviations otherwise.

Figure \ref{fig:charger_timeline} shows the times at which each charger is used during the day for both the integrated control architecture and the static baseline. Note that charging can only begin after 1 hour and 20 minutes of simulation time, as explained previously. It can be observed that the charging schedule is very tight in this simulation as both chargers are in use most of the time. It is interesting to note the difference in how charging is distributed throughout the day between the two methods. With the static baseline, charging events are uniformly distributed throughout the day regardless of the electricity prices, as can be seen in Figure \ref{fig:charger_timeline_baseline}. On the other hand, chargers are used less at times of higher electricity prices with the integrated control architecture, as symbolized by the gaps in the charger timelines from the 3rd to the 5th simulation hours and from the 12th to the 14th simulation hours in Figure \ref{fig:charger_timeline_controller}. Charging events are instead concentrated during periods when electricity prices are lower.

Figure \ref{fig:optimization} shows what the computational performance of the Lagrangian heuristic in the proposed integrated control architecture usually looks like for one simulation instance. The optimality gaps, as well as the computation times of the method, tend to fluctuate during a simulation. While these fluctuations usually vary from one simulation instance to the next, a few observations can be made consistently across most simulation instances. First, runtime values tend to be concentrated around 90 seconds here when using $n_\text{iter}=5$ for the subgradient algorithm, which is much below the update frequency of $5$ minutes of the high-level control layer. There are usually a few outliers where the runtime can be longer, up to 4 minutes in some instances. If these outliers are judged to be unacceptable because their computation times get too close to the update frequency of the references it is always possible to reduce the computation times to an acceptable value by decreasing the number of subgradient iterations on the fly. In addition, it can be observed in Figure \ref{fig:optimization} that the Lagrangian heuristic manages to solve several of the optimization problems to optimality. While this indicates that the duality gap is zero for at least some of the problems, this conclusion cannot be extended to the rest of the problems where optimality gaps of 5 to 10 $\%$ are routinely obtained with the Lagrangian heuristic. Finally, it was observed that a combination of zero optimality gaps and abnormally long runtimes was often obtained between the 11th and 13th simulation hours, something which is clearly visible in Figure \ref{fig:optimization}. We believe that this is the result of having vehicles with state-of-charge values close to $\sigma_\text{min}$ during evening rush hours: some buses cannot skip charging and must charge above $\sigma_\text{min}$, which results in the other buses having to skip charging in order to manage the increased load on the transit system. This might result in a reduced number of feasible solutions to the optimization problem \eqref{eq:MILP}. If such is the case, feasible solutions might be harder to find in the Lagrangian heuristic, thus causing longer runtimes, but any feasible solution is more likely to be the optimal solution, thus resulting in low or even zero optimality gaps.

\begin{table*}[!t]
\renewcommand{\arraystretch}{1.1}
\caption{Main result table. $\overline{p}_\text{el}$ denotes the average daily electricity price for each time-of-use pricing profile. The cost reduction is calculated between the proposed integrated control and the best-performing baseline for each category. The $CV^2$ values given are for bus routes 7, 12, and 85, respectively.}
\label{tb:results}
\centering
{\begin{tabular}{|l|l|c|c|c|c|c|} \hline
& & Service cost & Charging cost & Total cost & $CV^2$ & Conflicts \\ \hline
\textbf{2023-02-10} & Integrated control & $3328$ \euro & $277$ \euro & $3605$ \euro & $0.094$/$0.032$/$0.016$ & $0.68\%$ \\
$\overline{p}_\text{el}=44$ \euro & Adaptive baseline & $3769$ \euro & $330$ \euro & $4099$ \euro & $0.170$/$0.056$/$0.029$ & $9.59\%$ \\
& Static baseline & $3868$ \euro & $375$ \euro & $4243$ \euro & $0.170$/$0.059$/$0.027$ & $6.72\%$ \\ \cline{2-7}
& Cost reduction & $11.7\%$ & $16.1\%$ & $12.1\%$ & & \\ \hline \hline
\textbf{2022-02-10} & Integrated control & $3229$ \euro & $376$ \euro & $3605$ \euro & $0.071$/$0.037$/$0.014$ & $0.21\%$ \\
$\overline{p}_\text{el}=55$ \euro & Adaptive baseline & $3673$ \euro & $466$ \euro & $4139$ \euro & $0.111$/$0.046$/$0.027$ & $8.86\%$ \\
& Static baseline & $3868$ \euro & $473$ \euro & $4341$ \euro & $0.170$/$0.059$/$0.027$ & $6.72\%$ \\ \cline{2-7}
& Cost reduction & $12.1\%$ & $19.3\%$ & $12.9\%$ & & \\ \hline \hline
\textbf{2022-01-01} & Integrated control & $3093$ \euro & $500$ \euro & $3594$ \euro & $0.068$/$0.032$/$0.012$ & $0.24\%$ \\
$\overline{p}_\text{el}=82$ \euro & Adaptive baseline & $3824$ \euro & $623$ \euro & $4447$ \euro & $0.163$/$0.057$/$0.046$ & $6.88\%$ \\
& Static baseline & $3868$ \euro & $631$ \euro & $4499$ \euro & $0.170$/$0.059$/$0.027$ & $6.72\%$ \\ \cline{2-7}
& Cost reduction & $19.1\%$ & $19.7\%$ & $19.2\%$ & & \\ \hline \hline
\textbf{2022-09-14} & Integrated control & $3286$ \euro & $614$ \euro & $3900$ \euro & $0.072$/$0.038$/$0.012$ & $0.17\%$ \\
$\overline{p}_\text{el}=93$ \euro & Adaptive baseline & $3893$ \euro & $692$ \euro & $4585$ \euro & $0.107$/$0.101$/$0.032$ & $8.94\%$ \\
& Static baseline & $3868$ \euro & $808$ \euro & $4676$ \euro & $0.170$/$0.059$/$0.027$ & $6.72\%$ \\ \cline{2-7}
& Cost reduction & $15.0\%$ & $11.3\%$ & $14.9\%$  & & \\ \hline \hline
\textbf{2022-06-21} & Integrated control & $3524$ \euro & $1111$ \euro & $4635$ \euro & $0.068$/$0.036$/$0.012$ & $0.13\%$ \\
$\overline{p}_\text{el}=180$ \euro & Adaptive baseline & $3946$ \euro & $1260$ \euro & $5206$ \euro & $0.100$/$0.099$/$0.042$ & $8.19\%$ \\
& Static baseline & $3868$ \euro & $1449$ \euro & $5317$ \euro & $0.170$/$0.059$/$0.027$ & $6.72\%$ \\ \cline{2-7}
& Cost reduction & $8.9\%$ & $11.8\%$ & $11.0\%$ & &\\ \hline \hline
\textbf{2022-12-16} & Integrated control & $3590$ \euro & $2425$ \euro & $6015$ \euro & $0.071$/$0.046$/$0.015$ & $0.02\%$ \\
$\overline{p}_\text{el}=411$ \euro & Adaptive baseline & $3868$ \euro & $2933$ \euro & $6801$ \euro & $0.153$/$0.074$/$0.029$ & $6.66\%$ \\
& Static baseline & $3868$ \euro & $3140$ \euro & $7008$ \euro & $0.170$/$0.059$/$0.027$ & $6.72\%$ \\ \cline{2-7}
& Cost reduction & $7.2\%$ & $17.3\%$ & $11.6\%$ & & \\ \hline
\end{tabular}}
\end{table*}

The main results of the case study are now presented in Table \ref{tb:results}. For each time-of-use pricing profile, the values displayed in this table are averaged over all repetitions. As before, the time-of-use electricity pricing profiles are sorted in order of increasing mean electricity prices. The metrics of interest presented in this table include the service and charging costs, but also the squared coefficient of variation of headways $CV^2$ and the proportion of time spent waiting for a charger at the terminal. The service cost and charging cost are computed from their expressions as the first and third terms in the objective function \eqref{eq:objective_function}, respectively. The squared coefficient of variation of headways $CV^2$ is a performance indicator that measures headway regularity and which is often used to study the operational stability of high-frequency bus routes \citep{berrebi2}. Lower values of $CV^2$ denote regular headways whereas higher values indicate a high variance of the headways. Finally, the proportion of time waiting for a charger at the terminal is noted as “Conflicts” in Table \ref{tb:results} and is defined as the total amount of time by which vehicles must delay their charging plans when all chargers are used over the total amount of time spent by vehicles at the shared terminal.

It can be seen in Table \ref{tb:results} that the proposed integrated control architecture outperforms both baselines in terms of service regularity and charging cost across all the time-of-use pricing profiles tested. Compared to the best-performing baseline in each category, the proposed method achieves savings in service cost ranging from 7$\%$ to 19$\%$ and savings in charging cost ranging from 11$\%$ to 20$\%$, which results in overall cost savings of 11$\%$ to 19$\%$. Similarly, the $CV^2$ values of the integrated control architecture are lower for all bus routes, which means that it is better at keeping regular headways between buses than the baselines are. It can also be observed that the adaptive baseline consistently leads to smaller charging costs than the static baseline, as its charging decisions are taken based on current electricity prices. However, none of the baselines is clearly better than the other in terms of service regularity, judging from the service costs and the $CV^2$ values. Note that, in this table, the operational metrics have the same values for the static baseline for all time-of-use pricing profiles, only the charging cost values are different. This observation is not surprising since the static baseline relies on fixed charging times for each bus line and is therefore agnostic to electricity prices, and the same random seeds have been used to calibrate traffic simulations for all time-of-use pricing profiles. Therefore, the same control decisions are taken by the static baseline for all time-of-use pricing profiles. Note also that, judging from the $CV^2$ values presented in the table, bus route 7 seems to be more vulnerable to headway instability than the other routes since the corresponding $CV^2$ values tend to be higher.

When it comes to the proportion of time spent waiting for a charger, which is displayed in the last column of Table~\ref{tb:results}, the integrated control architecture clearly stands out from the baselines. Indeed, buses controlled with the integrated control architecture spend on average much less than 1$\%$ of their time at the terminal waiting for a charger to become available. By contrast, this figure is close to 7$\%$ for the static baseline, and usually around 7 to 9$\%$ for the adaptive baseline. What this means is that baseline buses are frequently affected by overlaps of their charging plans. It often happens that one bus has to postpone its planned charging period due to all chargers already being occupied. This is especially true for the adaptive baseline, which consistently has more charging conflicts than the other methods, as can be seen in Table \ref{tb:results}. The reason for this is that vehicles are incentivized to charge their batteries when electricity is cheaper with the adaptive baseline, but without any coordination between them. Consequently, the charging demand increases during hours with lower electricity prices and far exceeds the charging capacity of the system, which greatly increases the number of charging conflicts. The conflict proportion is smaller for the static baseline since the charging demand is distributed uniformly throughout the day instead of being concentrated at specific hours with a lot of overlaps, even though here too there is no coordination between vehicles when accessing the chargers. This is precisely why the integrated control architecture manages to stand out on this performance metric. Thanks to its predictive feature, the integrated control architecture can anticipate potential charging conflicts and coordinate bus charging plans so as to avoid any overlap. This ensures that chargers are fully utilized during hours with low electricity prices, thus resulting in lower charging costs. But more importantly, avoiding charging conflicts means that vehicles are more rarely idling at the terminal waiting for a charger. With the integrated control architecture, these undesired holding times are instead converted to additional slack added to the transit system. This makes the transit system more resilient to external disturbances, such as the dense road traffic of the rush hours, and improves the ability of vehicles to maintain regular headways. This in turn explains the better performances observed for the proposed integrated control architecture in terms of service cost and $CV^2$ values.

It can be observed in Table \ref{tb:results} that the service cost of the integrated control architecture is higher for the 2022-06-21 and the 2022-12-16 time-of-use pricing profiles, which are the ones with the highest average electricity prices. This results in a lower service cost reduction compared with the baselines. We believe that this effect is primarily due to the fact that the higher electricity prices lead to stronger penalties in the objective function \eqref{eq:objective_function} when charging at times of high electricity prices. These strong penalties incentivize the integrated control architecture to carry out bus charging at other times, even when it comes at the expense of a degraded level of service, e.g. by making buses charge longer when prices are low and dispatching them behind schedule. In other words, the trade-off between low charging costs and low service costs in the objective function tends to move in the direction of the former when electricity prices are high on average, which may result in increased service costs. On the other hand, no clear pattern appears for the charging cost. Of course, charging costs increase for all methods with the increase of the average daily electricity price, but the charging cost reduction seems to fluctuate randomly between 11$\%$ and 20$\%$ for all time-of-use pricing profiles studied.

To summarize, one key finding is that both the integrated control architecture and the adaptive baseline consistently manage to obtain lower charging costs compared to the price-agnostic static baseline. Thus, it can be concluded from this case study that prior knowledge of hourly electricity prices can be leveraged by transit agencies to adapt the charging schedule of their vehicles and reduce their charging costs. As we have seen, this does not necessarily lead to a degradation in the level of service, even in cases like here where the charging schedule is very tight. Another key finding from this case study is that including operational aspects when taking charging scheduling decisions, as is done in the proposed integrated control architecture but not in the two baselines considered, can lead to cost reductions both in terms of service-related costs and charging costs. We think that this is especially true when the shared charging capacity is limited, as in this case study, since a lack of coordination in the operational control of vehicles can result in an increased number of conflicts when accessing the chargers and thus disrupt the regularity of bus service. Running an optimization-based procedure like the proposed integrated control architecture requires to have an adequate computing and communication infrastructure, however, and it is ultimately up to transit agencies to decide whether the expected cost savings are worth investing in the equipment necessary to run such methods.

\section{Conclusion} \label{sec:6}

In this paper, a hierarchical control framework has been proposed to solve the integrated operational control and charging scheduling problem for electric bus networks relying on daytime opportunity charging at terminals or bus depots. By separating high-level travel time and charging decisions from the low-level control of each vehicle, bus operation can be predicted and controlled over horizons of several hours. This makes it possible for vehicles to adapt their use of the charging infrastructure to the current operational state of the transit system, which has been found to be of particular importance when the charging capacity is limited and shared by several bus lines. The integrated control problem is formulated as an MILP in this work, but this optimization problem has been found to be too large to be solved efficiently with off-the-shelf solvers in most practical situations. Instead, a Lagrangian heuristic combining Lagrangian relaxation with a local search heuristic is proposed to solve it. The charger exclusion constraints, which restrict the number of vehicles that can access the chargers simultaneously, are the only source of coupling between different bus lines and are therefore relaxed to form the Lagrangian dual problem. The dual problem is in turn separable and can be decomposed into independent MILPs of much smaller size. This feature can be exploited when deploying the subgradient algorithm to solve the Lagrangian dual problem by solving these subproblems in parallel. The proposed local search heuristic works by taking some of the infeasible solutions generated during the subgradient iterations and by reorganizing the planned charging events so as to remove any overlap.

The integrated control architecture developed in this work is tested both in synthetic simulations, in order to estimate how well the Lagrangian heuristic scales with the size of the optimization problem, and on a realistic case study based on historical data from the city of Chicago. It was found in the synthetic simulations that the Lagrangian heuristic could achieve reasonable optimality gaps in just a few subgradient iterations and with acceptable computation times for large bus networks of up to 20 bus lines. By contrast, it takes unreasonably long computation times for an off-the-shelf solver like Gurobi to obtain anything but large optimality gaps even on the smaller networks tested, which makes this approach impossible to use in practice. In the Chicago case study, the proposed integrated control architecture is compared to two rule-based control baselines which we think reflect current transit agency practices. The use of the microscopic traffic simulator Vissim made it possible to run high-fidelity simulations where the influence on bus service of such features as passenger accumulation at stops or vehicle queues at traffic lights could be captured explicitly. It was found in the case study that the integrated control architecture is better than the baselines at keeping a regular level of service and at reducing charging costs. This success is mainly explained by the fact that buses spend a lot less time waiting for a charger in case of conflicts in the charging schedule with the proposed method. All the undesired holding time resulting from overlaps in vehicle charging plans with the baselines is instead converted to additional slack added to the transit system with the integrated control architecture, which makes it more robust to disturbances in service regularity. In addition, the results from the case study demonstrate the value of knowing the hourly electricity prices in advance for charging scheduling. Control methods that had access to this information in the case study successfully managed to reduce their charging costs by adapting the charging schedule to the electricity prices, insofar as it did not affect the level of service too negatively.

We think that the next step to improve the work presented in this article would be to adapt the proposed formulation of the operational bus network control and charging scheduling problem to transit system models where some of the bus routes overlap, in order to model real life bus networks more faithfully. This would require more detailed passenger models, since some of the passengers would now have to choose between buses from several different bus lines, and these models would likely add additional coupling between bus lines in the overall problem. An interesting direction for future work would then be to study this new coupling structure and see if the Lagrangian relaxation and dual decomposition procedure presented in this paper could still be applied. Another ambitious future research direction would be to extend the proposed control framework to integrate another aspect of the electric bus network planning process. For instance, the vehicle scheduling problem could in principle be treated jointly with the charging scheduling problem and the real-time operational control problem, although more research will be needed to tackle the complexity of the resulting problem.

\section*{Acknowledgment}

The authors are very grateful to Maxime Sechehaye for setting up and calibrating in Vissim the simulation environment that
was used for the case study.




\bibliographystyle{cas-model2-names}
\bibliography{refs}







\end{document}